\documentstyle[aaspp4,psfig]{article}
\begin{document}
\title{ Cherenkov-Curvature Radiation and  Pulsar Radio Emission Generation}
\author{Maxim  Lyutikov }
\affil{Theoretical Astrophysics, California Institute of Technology,
Pasadena, California 91125}
\author{George Machabeli}
\affil{Abastumani Astrophysics Observatory, A. Kazbegi Av. 2a, Tbilisi,
380060
 Republic of Georgia}
\author{Roger Blandford}
\affil{Theoretical Astrophysics, California Institute of Technology,
Pasadena, California 91125}

\date {draft}

\begin{abstract}
Electromagnetic processes associated with a 
  charged particle moving in a strong 
circular magnetic field are considered in cylindrical coordinates.
We investigate  the relation between the vacuum curvature
emission and  Cherenkov emission and argue that, for the superluminal
motion of a particle in the inhomogeneous magnetic field
in a dielectric, the combined effects of magnetic field
inhomogeneity and the presence of a medium give rise to the synergetic
Cherenkov-curvature emission process. We find 
the conditions  when the operator relations between electric field
and electric displacement in cylindrical coordinates
may be approximated by  algebraic relations.
For  nonresonant
  electromagnetic wave, the 
interaction with  particles  streaming along
the curved magnetic field  may be described in the WKB approximation.
For resonant waves, interacting with superluminal particles
we use a plane wave approximation to compute
 the local dielectric tensor
of a plasma in a weakly inhomogeneous magnetic field.
We find in this approximation the polarization of normal modes in the plasma,
Cherenkov-curvature and Cherenkov-drift  emissivities  and growth
rates. 
\end{abstract}


\section{Introduction}

Despite much theoretical effort, there is still no widely 
accepted explanation of how pulsars emit the high
brightness radio emission by which they were first discovered
over thirty years ago. Difficulties with "antenna" mechanisms
 have led to
a resurgence of interest in "maser" mechanisms where an inverted
population of electrons and/or positrons amplifies an outgoing 
wave mode (e.g. \cite{Melrose-DB}). Two such
maser process that are particular promising are  the
Cherenkov-drift mechanism (which is Cherenkov-curvature 
emission with a drift (\cite{LyutikovMachabeliBlandford1})) and
cyclotron-Cherenkov emission at the anomalous Doppler resonance
(\cite{Kazbegi}).
This paper is concerned with elucidating the 
physics of the  Cherenkov-curvature and 
Cherenkov-drift  processes  and developing a new
mathematical description of it using cylindrical
coordinates. 
Using this approach, its close relationship  to the pure Cherenkov
emission processes becomes clear.

Pulsar radio emission is believed to originate on the open magnetic 
field lines that trace a path from the neutron star surface to the light 
cylinder at $r = c/\Omega$ and beyond. The field geometry is complex close
to the star at $ r  \approx R_{\ast} \approx 10 $ km and near the light
cylinder. However, it should be primarily dipolar for $R_{\ast} \ll
r  \ll c/\Omega$. The characteristic length scale, essentially the radius
of curvature of the field lines, $ R_B \approx (c r /\Omega)^{1/2}$,
is very large compared  with the wavelength of interest,
$\lambda$, and so we should be able to adopt a WKB approach following 
following individual wave packets as they propagate out through
the magnetosphere. However, in order to compute the 
local emission and absorption it is necessary
that the magnetic field remains curved. 
A simple model problem, that retains the essential
ingredients, comprises  a set of circular concentric cylindrical 
 magnetic  surfaces.  
 (Fig. \ref{max1}). We show, that it is possible to 
 ignore the radial variations in the strength
of ${\bf B}$ in computing the local interaction, provided that
this variation is not very rapid. In this model
problem the plasma
circulates continually along the circular
trajectories.

It is generally presumed, that electrons and positrons are in their
ground gyrational state and follow the curved field lines. This is 
a good approximation for the bulk of the plasma, which we will
suppose travels with a Lorentz factor $\gamma_p \approx 100$.
However, we believe that there is also a population of 
ultra energetic particles with $\gamma =10^{5-7}$ 
and, in the outer magnetosphere, these will experience a curvature
drift  relative to the bulk plasma. This has the
interesting consequence that their main electrodynamic interaction
is with the waves propagating in the bulk plasma at a finite 
angle to the magnetic field ${\bf B}$. It is these waves, that, we assert,
become the high brightness electromagnetic waves that escape from the 
magnetosphere. 

When considering the wave amplification in this problem, 
we assume that  fluctuating currents, present on the field
lines  at radii smaller than the radius of the considered region, 
produce electromagnetic fluctuations that are subsequently
amplified  due to the interaction with 
resonant particles  (Fig. \ref{max1}).
This is   a convective  type
instability, when a wave is amplified as it propagates
through an active medium.

A first attack on this problem was made by \cite{Blandford1975}
In this paper
the energy transfer from a plane infinite electromagnetic
wave to a single electron moving with
ultrarelativistic speed along a curved trajectory was found to be
always positive, independent of the distribution function and so there is no
 possibility of wave growth. 
The subsequent works (\cite{ZheleznyakovShaposhnikov}, 
\cite{MelroseLou},  \cite{Melrosebook1}) basically followed this
approach, which  emphasizes the 
analogy between  curvature emission and  conventional cyclotron emission. 
This approach, though formally correct, has  limited 
applicability and  ignores  two  important 
features of the emission mechanism.  The first is that, in adopting 
a plane wave formalism, the interaction length for an individual
electron,  $\approx R_B /\gamma_b$, 
was essentially coextensive with the  region over which the waves could
interact with any electron. This approach precludes a  strong amplification
under all circumstances because the wave would have to grow substantially
during a single interaction. The second problem was that a dispersion of the waves
was neglected.
We address the first shortcoming by expanding the electromagnetic 
field in cylindrical waves centered on $r=0$,
 and the second
explicitly by considering general plasma modes. 

In a separate approach developed by \cite{beskin1} 
an attempt was made to incorporate the collective effects 
of a plasma in an inhomogeneous magnetic 
 field. The fundamentals of that approach have been seriously
criticized  (\cite{Nambu}, \cite{machabeli}).

One of the key new elements in our approach is that we  expand the electromagnetic
fields in cylindrical coordinates and consider
 {\it resonant } interaction between a particle and these modes.
A common procedure in calculating the  energy emitted  by a given 
current distribution is to find the power emitted into a given  normal mode
of the medium and sum over all the modes. 
A power emitted into a normal  mode  by a given charge
distribution is  proportional to the square of the expansion of this current
in normal modes.
 In  a homogeneous   or weakly
inhomogeneous stratified medium considered in Cartesian 
coordinates the normal modes are plane waves 
(with slowly changing parameters), 
so that the power emitted into a given normal mode turns out to be proportional
to the Fourier transform of the current.
To find a similar expression for a single particle emissivity
into a cylindrical mode in an inhomogeneous medium is difficult  problem
for two reasons: there is a complicated, radius dependent,
form of the vector cylindrical waves 
and the dielectric response in an 
 inhomogeneous medium is nonlocal.
We address these complications  by finding the conditions, when 
the interaction of a particle with cylindrical waves can be 
approximated by an   interaction of a particle with plane waves.
We find two cases when this
can be done.
 First, this may be a true approximation 
for the{\it  nonresonant}  modes. This is equivalent to the 
 WKB approximation to the radial dependence
of  normal modes (mathematically, this corresponds
 to the tangent expansion of Bessel functions
when argument is significantly larger  that he order). In this case the
response of a medium becomes local.
Second, we find a particular case, when the 
{\it resonant} modes can be approximated as local plane waves.
 A resonant interaction of a relativistic particle
with a cylindrical mode occurs near the point when  the  argument of Bessel functions
is close to the order. The WKB approximation, or expansion
in tangents,  is not applicable in this case and we have to use the
 Airy function
approximation to Bessel function, which has a plane wave approximation
for the interaction of {\it subluminous} waves with the  particles
moving  with  speed larger than the  speed of light in a medium 
This corresponds to the Airy function expansion
argument being larger than the order.

We should note  here, that 
the electrodynamics of the interaction of a particle moving 
along the curved magnetic field with
speed larger than the speed of light in the medium
 is quite unusual and can be considered
as a new, Cherenkov-Curvature, emission mechanism 
(\cite{LyutikovMachabeliBlandford1}),
 which differs from  conventional Cherenkov, cyclotron or curvature
emission and includes, to some extent, the features of each of these mechanisms.
In an extension of ideas of  \cite{Schwinger},
  conventional synchrotron emission
and Cherenkov radiation may be 
 regarded as respective limiting cases
 of $|n-1|\, \rightarrow \,1 $ and $B \rightarrow \,0$
of a  "synergetic" cyclotron-Cherenkov radiation, 
the Cherenkov-curvature radiation necessary
includes  effects of the magnetic field gradients.

When the dielectric response of a medium
to cylindrical wave is {\it local}
it is possible to calculate a simplified dielectric tensor. 
Using this dielectric tensor we find the normal modes of  strongly
magnetized electron-positron plasma and show that a beam of particles
propagating along the curved magnetic field can amplify the electromagnetic
waves. The amplification occurs at the
Cherenkov-drift resonance $\omega-k_{\phi} v_{\phi} - k_x u_d=0$
($\omega$ is the frequency of the wave, $k_{\phi}$, $k_x$ and $v_{\phi}$
 are 
the corresponding projection of the
wave vector and velocity and $u_d$ is a curvature drift velocity).
The fact that this is a Cherenkov-type resonance immediately 
implies that  the presence of a subluminous waves with the
phase velocity smaller that the speed of light is essential.
The presence of a drift provides a coupling between the
the electric field of the electromagnetic waves  
 particle's motion.
Our estimates show, that this instability can grow fast enough
to account for the observed pulsar radio emission.
This instability may be regarded as new type of a curvature maser. 
An interesting and peculiar feature of this mechanism, is that,
unlike  with conventional curvature emission,
the emitted waves have a polarization almost 
perpendicular to the osculating
plane of the magnetic field.

The overview of this work is the following.
In Section 
\ref{vacum} we analyze the properties of electromagnetic
waves in cylindrical coordinates in vacuum and the
interaction of the vacuum waves with a charged particle.
 In Section \ref{SingleParticle} we analyze  in cylindrical coordinates
the 
electromagnetic fields of a particle moving along a spiral
trajectory using   the dyadic Green's function  for the vector wave 
equation, derive the emissivity of a particle in ground
gyrational level into a cylindrical mode and
 rederive the curvature emissivity in cylindrical coordinates.
 This is followed in Section 
\ref{Wavesisotroipicdielectric} by the generalization to a dispersive isotropic 
medium and the importance 
of waves with the phase speed less than the speed of light in vacuum is brought out.
In Section \ref{DispersionInf} we consider {\it anisotropic}
plasma in infinitely strong magnetic field and derive the 
curvature emissivity using the Vlasov approach. 
In Section \ref{Airyfunction} we discuss the various regimes
of the Airy function approximation to the Cherenkov-curvature
emission and find the conditions for the plane-wave approximation. 
In Section \ref{WavesAsymptotic} we investigate the features of the electromagnetic waves
in the asymptotic regime $z \gg 1$ and 
in Section \ref{ResponceTensor} we calculate the response tensor
for  a one dimensional plasma in a strong curved magnetic
field taking into account the drift velocity. 
Finally, in Section \ref{pl} we analyze the polarization
properties of electromagnetic waves in the plane-wave approximation
and calculate the increment of the Cherenkov-drift instability.

\section{Vacuum Solutions}
\label{vacum}

\subsection{Vacuum Normal Modes}
\label{VacuumNormalModes}

We expand 
the fundamental  solutions of the wave equation 
\begin{equation}
{\bf curl \, curl\,  E } - {\omega^2 \over c^2} {\bf E } =0
\label{D1}
\end{equation}
in terms of Fourier amplitudes in $x$ and $\phi$ coordinates and time:
\begin{equation}
{\bf E} ({\bf r} , t)  =
\sum_{\nu  = - \infty} ^{\nu   = \infty}  \int d \omega  \int {d k_x \over 2 \pi} 
{\bf E }
(r, k_x,\nu , \omega)
\exp \left\{- i(\omega t -\nu \phi - k_x x) \right\}.
\label{D11}
\end{equation}
The wave equation (\ref{D1}) then takes the form
\begin{eqnarray}
&&
{i\nu \over r^2}  {\partial  \over \partial r} 
\left(r E_{\phi} \right) +
i k_x  {\partial \over \partial r} E_x  -{k_x\nu \over r} E_{\phi} +
\left({\nu^2 \over r^2} - \left({\omega^2 \over c^2 } -k_x^2 \right) \right)
E_r=0
\label{D1400}
\mbox{} \\ \mbox{}
&&
-
{\partial \over \partial r} \left({1\over r}  {\partial \over \partial r} 
\left(r E_{\phi}  \right)  \right) +  i\nu {\partial \over \partial r} 
\left({E_r \over r } \right)  -
{k_x\nu \over r} E_x - \left({\omega^2 \over c^2 }  -k_x^2 \right) 
 E_{\phi}  =0
\label{D14001}
\mbox{} \\ \mbox{}
&&
-  {1\over r} {\partial \over \partial r}  
 \left(r {\partial \over \partial r} 
E_x  \right) +
{i k_x \over r} {\partial \over \partial r} 
\left(r E_r \right) -
\left({\omega^2 \over c^2 } - {\nu^2 \over r^2} \right) E_x =0
\label{D14002}
\end{eqnarray}
Using the fact that
in a wave the
magnetic field is related to the electric field:
\begin{equation}
{\bf curl \, E} = -{1\over c} {\partial {\bf B} \over \partial t} =
i{\omega \over c}  {\bf B}.
\label{D14012}
\end{equation}
we find
the amplitudes $ {\bf E }
(r, k_x,\nu , \omega)$, subject to condition of being finite at $r=0$
\begin{eqnarray}
&&
{\bf E } ^{(lt)} (r, k_x,\nu , \omega) = E^{(lt)}  
\left({i k_x \over \lambda} J_{\nu }   ^{\prime} (\lambda r )  {\bf e_r}-
{\nu  
k_x \over \lambda^2 r} J_{\nu } (\lambda r ) {\bf e_{\phi}} + J_{\nu } (\lambda r )
 {\bf e_x} 
\right)
\label{D02}
\mbox{}  \\ \mbox{}
&&
{\bf B } ^{(lt)} (r, k_x,\nu , \omega) = {\omega  \over \lambda^2 c } E^{(lt)}
\left({\nu J_{\nu } (\lambda r )\over r}  {\bf e_r} + i  \lambda
J_{\nu } ^{\prime} (\lambda r ) \, {\bf e_{\phi}} \right)
\label{D2}
\mbox{} \\ \mbox{}
&&
{\bf E } ^{(t)} (r, k_x,\nu , \omega)   =  E^{(t)}  \,
\left({i\nu \over  \lambda r } J_{\nu } (\lambda r ) {\bf e_r} - 
 J_{\nu }^{\prime}  (\lambda r ) {\bf e_{\phi}} 
\right)
\label{D03}
\mbox{}  \\ \mbox{}
&&
{\bf B } ^{(t)} (r, k_x,\nu , \omega) =  E^{(t)} 
\left({k_x c  \over \omega}    J_{\nu }  ^{\prime} (\lambda r ) {\bf e_r} +
{i k_x c\nu  \over  \lambda r \omega }  J_{\nu } (\lambda r )  {\bf e_{\phi}}
- i { c 
\lambda \over \omega}   J_{\nu } (\lambda r ) {\bf e_x} \right)
\label{D3}
\end{eqnarray}
where $ \lambda = \sqrt{\omega^2/c^2 - k_x^2} $,
$J_{\nu } (\lambda r )$ are Bessel functions,
which satisfy the equation
\begin{equation}
{1\over r} {1\over \partial r} \left( r {1\over \partial r} J_{\nu}\right) -
\left({\nu^2\over r^2} + k_x^2 -{\omega^2 \over c^2} \right) J_{\nu}=0,
\label{D702}
\end{equation}
 and 
$ {\bf e_r}, {\bf e_{\phi}} $ and  ${\bf e_x}$ are unit vectors
along the corresponding axes, $E^{(lt)}$ and ${\bf E } ^{(t)}$ are amplitudes
of the modes. The superscripts denote polarization: ${\bf E } ^{(lt)}$ is 
perpendicular to the coordinate surface ${\bf e_x}$
and ${\bf E } ^{(lt)}$ lies on the coordinate surface ${\bf e_x}$.

\subsubsection{Short Radial Wave Length Expansion of Vacuum Solutions}
\label{ShortWaveLength}

For large orders $ \nu \gg 1$, argument larger than the
order  $\lambda r > \nu $  and not very close to it,
$\lambda r -  \nu \geq \nu $, we can use expansion in tangents
of the  Bessel functions: 
\begin{equation}
 J_{\nu } (\nu \sec \zeta ) \approx \sqrt{{2\over \pi  \nu \tan \zeta }}
\cos  \left(\nu (\tan \zeta -  \zeta ) + {\pi \over 4} \right),
\hskip .2 truein 
\mbox{$ \zeta \geq \nu ^{-1/2}  $  }
\label{D71}
\end{equation}
In our case $ \cos \zeta = \nu /(\lambda r)$.

Introducing 
\begin{equation}
k_r^2 = \lambda^2 -{\nu^2 \over r^2},
\label{D710}
\end{equation}
the  Eq.   (\ref{D71}) for the Bessel function
may be viewed as a dispersion relation:
\begin{equation}
k_r^2 + k_{\phi}^2 +k_x^2 = {\omega^2 \over c^2}
\label{D72}
\end{equation}
Then the expansion (\ref{D71}) reads
\begin{equation}
 J_{\nu } (\lambda r) \approx
\sqrt{{2\over \pi  k_r r }} \cos \left(\pm  k_r r - \phi_{\nu} \right)
\label{D711}
\end{equation}
where $\phi_{\nu}$ is insignificant phase shift.
The limits of applicability of this expansion is that $ k_r r \gg 1$ and 
$\zeta$ not very small.
The normal modes in this limit
are 
\begin{eqnarray}
&&
{\bf E } ^{(lt)} ({\bf r}, t) \propto 
e^{- i (\omega t -k_x x \pm  k_r  r )}
\left(- {k_x k_r \over \lambda^2} {\bf e} _r - {k_x k_{\phi} \over \lambda^2}  
{\bf e} _{\phi} + {\bf e} _x \right) 
\mbox{} \nonumber \\ \mbox{}
&&
{\bf E } ^{(t)} ({\bf r}, t) \propto
e^{- i (\omega t -k_x x \pm   k_r  r ) }
\left({ k_{\phi} \over \lambda}
 {\bf e} _r -  { k_r  \over \lambda} {\bf e} _{\phi}  \right)
\label{D721}
\end{eqnarray}

\subsubsection{WKB Solution (Short Radial Wave Length)}
\label{WKBSolution}

We can obtain solutions  (\ref{D721}) of Eqns (\ref{D1400}- \ref{D14002})
using  WKB procedure, when the effective radial wavelength is much shorter
than the characteristic scale of the problem.
Assuming that the solutions
of  Eqns (\ref{D1400}- \ref{D14002}) have a form
\begin{equation}
{\bf E }  \propto
{1\over \sqrt{r}} e^{i S},
\hskip .2 truein 
{\partial S \over \partial r } \gg {1 \over r }
\label{D725}
\end{equation}
we find 
\begin{equation}
S = \pm \int dr  \sqrt{\lambda^2 - {\nu^2 \over r^2} } =
\pm \left(k_r r - \nu \arccos { \nu \over \lambda r } \right)
\label{D726}
\end{equation}
with $k_r $ defined by Eq. (\ref{D72}).

The dispersion equation, which is obtained by substituting Eq. (\ref{D725}) in
Eqns (\ref{D1400}- \ref{D14002}), takes a form 
\begin{equation}
\epsilon^{(0)}_{ij} E_i =0
\label{D724a}
\end{equation}
where
\begin{equation}
\epsilon^{(0)}_{ij}=
\left(\begin{array}{ccc}
k_{\phi}^2 +k_x^2 - {\omega^2 \over c^2} & 
 -k_r k_{\phi} +{i k_{\phi} \over 2 r }  & 
-k_x k_r - {i k_x  \over 2 r } \\
-k_r k_{\phi} -{ 3  i k_{\phi} \over 2 r }  &
k_x^2 + k_{r}^2 + {3\over 4 r^2}  - {\omega^2 \over c^2} &
- k_x  k_{\phi}\\
-k_x k_r+  {i k_x  \over 2 r } &
- k_x  k_{\phi}&
k_{\phi}^2  + k_{r}^2 - {1 \over 4 r^2}  - {\omega^2 \over c^2}
\end{array}
\right)
\label{D72511}
\end{equation}

For $k_r \gg 1/r $ we can drop the complex part 
of the tensor $\epsilon^{(0)}_{ij}$ (which is not even Hermitian 
since $k_r$ is not a Killing vector).
The dispersion equation det$||\epsilon^{(0)}_{ij} ||=0$ then
gives  again
\begin{equation}
k_r^2 +k_x^2 +k_{\phi}^2 =  {\omega^2 \over c^2}
\label{D726a}
\end{equation}

The Bessel equation (\ref{D702}) may be viewed as a Schr\"{o}dinger
type equation 
 for a wave functions of  a particle moving in a two dimensional potential
given by the $ \nu^2 /r^2 $ term in Eq. (\ref{D702}).
The characteristic scale of the  Schr\"{o}dinger-type equation (\ref{D702})
is $ r_{\nu} = \nu/ \lambda $. A point $r=r_{\nu} $ may be viewed as a
classical reflection point.
Solutions are exponentially decaying for smaller radii and
have a wavelike structure at larger radii.
 The analogy between the Eq. (\ref{D702} extends even further.
When approaching the  classical reflection point a common approach
in quantum mechanics is to use the Airy function approximation
to the solutions of the Schr\"{o}dinger equation.  This corresponds to the
transition region in the Bessel function expansion, when 
relations (\ref{D71}) are no longer valid and one should use 
Airy function expansion.
We will discuss the various regimes of the Airy function expansion
in Section \ref{Airyfunction}.

The WKB method breaks down when the radial wavelength becomes
comparable to the inhomogeneity scale of the "potential".
For a given radius this occurs for 
\begin{equation}
\nu \approx {\omega r \over c } \approx {r  \over \bar{\lambda}} 
= 10^8
\label{D727}
\end{equation}
for $r=10^9$ cm and the wavelength $  \bar{\lambda} =10$ cm.

The  break  down of the  WKB method may be illustrated graphically
using a different analogy. 
On a two-dimensional plane  
consider the surfaces of a constant  phase of Hankel function
$H_{\nu}^{(1)}$:
\begin{equation}
\phi = \arctan{\left({Y_{\nu} (\lambda r )
 \over J_{\nu} (\lambda r)} + {2 \pi l \over \nu }
\right) }
, \hskip .2 truein l = 0,1,2,... \nu-1
\label{D728}
\end{equation}
(Fig \ref{phase}).
This whole picture is rotating with a frequency $\omega$.
The points when the argument equals order, i.e., the "classical reflection
point", may be regarded as light cylinder.
For  radii much larger than the "light cylinder" radius,
$r_{\nu}
 = \nu/\lambda$, 
the surface of a constant phase has a form of  unwinding  spiral with
a wavelength $\lambda /\nu$. Near the light cylinder the radial wave
length becomes comparable to the light cylinder  radius. 

From the quantum mechanical point of view, the kinetic energy, given by the
derivative term in Eq . (\ref{D702}) becomes zero at the 
light cylinder radius. This is the classical
reflection
point. For larger radii the kinetic energy is real and positive:
the solution of the
 "Schr\"{o}dinger equation" (\ref{D702}) has a wavelike structure.
 For smaller  radii the kinetic energy is complex:
the solution of the "Schr\"{o}dinger equation" (\ref{D702}) has 
exponentially decaying form, modified by the presence of the 
of the reflection point at $ r = 0$.

\subsection{Wave-particle interaction}
\label{Wave-particle}

A considerable simplification of the wave-particle interaction
in cylindrical coordinates can be made when the  interaction 
occurs in  the  WKB limit, or, equivalently,
when we can use the expansion in tangent for the Bessel
functions. From Eq. (\ref{D71})  we find a lower limit on the 
azimuthal wave number:
\begin{equation}
\nu \geq (\lambda r )^{2/3} \approx 10^5
\label{D0a71}
\end{equation}
for the observed wavelength $10$ cm and the curvature radius $10^9$ cm.
From Eqs. (\ref{D727}) and (\ref{D0a71})
it follows, that  for a given radius and given wavelength
cylindrical waves with azimuthal wave numbers
in the range  $ (\lambda r )^{2/3} \leq  (\lambda r ) $
can be considered in the WKB limit, i.e. in the plane wave approximation.

It is  possible to picture graphically the resonant and nonresonant
 interaction
of a particle  moving along the circular trajectory.  
The pattern in Fig. \ref{phase} may be thought of as rotating
 with a frequency $\lambda c / \nu$. A particle at a given distance $r_0$
with a given velocity $v_{\phi}$ is rotating with an angular
frequency $\Omega= v_{\phi} /r_0$. A resonant interaction occurs
when these two frequencies are equal. In vacuum, 
\begin{equation}
r_0 = \nu \beta_{\phi} \lambda \approx  r_{\nu}\left(1-{1\over 2 \gamma^2} 
\right)
\label{D731}
\end{equation}
where $  r_{\nu} = \nu /\lambda$ is the light cylinder radius for the
mode $\nu$.
So, in vacuum, the resonance between a particle and a wave
always happens inside the light cylinder.

\section{Synchrotron Emission Considered in Cylindrical Coordinates}
\label{SingleParticle}

{ 
In this section we find  electromagnetic 
fields from a particle executing helical motion in vacuum and 
 rederive the conventional expressions for the
cyclotron (or curvature) emissivities using cylindrical 
coordinates. The results of this section shows
 how the conventional expressions for the
synchrotron emissivity in vacuum can be obtained using our 
approach.

We first find  the expansion of  electromagnetic fields due to a particle
executing helical motion in vacuum and in a homogeneous dielectric
  in terms of normal  vector modes
in cylindrical coordinates. This is done using  dyadic Green's
function that gives a response of a vector field to a vector source.
 We then show how  conventional 
expressions for synchrotron or curvature
emissivities can be obtained by two different
methods: first by calculating an emissivity into a given cylindrical mode using
an expansion of a current in terms of the normal vector modes
and secondly by calculating Poyting flux through cylindrical surface.
These calculations show how the resonant  wave-particle interaction
can be reformulated 
for  cylindrical waves for a particular choice of helical motion .
 
}

\subsection{Dyadic Green's Function}
\label{DiadicGreen}

In order to find the electromagnetic fields produced by a given charge
distribution in a medium
is it necessary to know the Green's function for the vector
d'Alambert's equation:
\begin{equation}
\Box {\bf F} ({\bf r}, t) = Q ({\bf r}, t) 
\label{D7291}
\end{equation}
The solution of the Eq.  (\ref{D7291}) may be written using {\it  dyadic}
Green's function ${\cal{G}} ( {\bf r},  t, {\bf r}^{\prime}  ,  t^{\prime})$
\begin{equation}
F ({\bf r}, t) = \int d {\bf r}^{\prime}  d t^{\prime}   {\cal{G}}
( {\bf r},  t, {\bf r}^{\prime}  ,  t^{\prime}) \cdot Q ({\bf r}^{\prime} , t^{\prime}  )
\label{D72911}
\end{equation}

In what follows we limit ourselves to the sources  changing with time 
$\propto e ^{ i \omega t}$. Then in vacuum d'Alambert's equation
reduces to Helmholtz equation
\begin{equation}
{\bf curl \, curl\,} {\bf F} ({\bf r},\omega) - {\omega^2 \over c^2} {\bf F} ({\bf r},\omega) =
 {\bf Q}({\bf r},\omega) 
\label{D729}
\end{equation}
where $ {\bf F}$ is electric or magnetic field and 
$ {\bf Q}$ is  a source function.
For a single particle source, which is a delta function in 
coordinates, Eq. (\ref{D729}) becomes an equation for the 
Green's function.
To find Green's function we first have to find the 
solutions to the homogeneous equation with $ {\bf Q}=0$.
The solutions to the homogeneous  vector wave equation can be obtained from the
solution of the scalar wave equation
\begin{eqnarray}
&&
{\bf curl \, curl\,} \Psi  - {\omega^2 \over c^2}  \Psi =0
\label{D3011}
\mbox{}  \\ \mbox{}
&&
\Psi(\nu , k_x, \lambda) =  e^{  i (  \nu  \phi + k_x x) }
 Z_{\nu }(\lambda r)
\label{D311}
\end{eqnarray}
($ Z_{\nu }$ is any Bessel function)
using the relations
\begin{equation}
{\bf E } ^{(t)} = {\bf curl}({\bf e}_x   \Psi) ,
\hskip .3 truein
{\bf E } ^{(lt)} = {\bf curl \, curl\,}  ({\bf e}_x \Psi)
\label{D312}
\end{equation}
Functions  (\ref{D311})
are  eigenfunction for the scalar wave equation.
The eigenfunctions for the vector  wave equation (\ref{D3011})
are (e.g. Eq. (6.5-6.6) of \cite{Tai}) 
\begin{eqnarray}
&&
{\bf L} ={1\over \lambda}  {\bf grad} \Psi  =
 e^{  i (  \nu  \phi + k_x x) }
\left(Z_{\nu }^{\prime} {\bf e}_r + {i \nu  \over  \lambda r} Z_{\nu }
{\bf e}_{\phi}+
 + {i k_x \over  \lambda} Z_{\nu }  {\bf e}_x\right)
\label{D031}
\mbox{}  \\ \mbox{}
&&
{\bf M} = {1\over \lambda}  {\bf curl} ({\bf  e}_x \Psi) 
 = 
 e^{  i (  \nu  \phi + k_x x) }
\left(i {\nu  \over  \lambda r} Z_{\nu }  {\bf e}_r -
Z_{\nu }^{\prime} {\bf e}_{\phi} \right)
\label{D033}
\mbox{}  \\ \mbox{}
&&
{\bf N} =  {c \over \omega  \lambda}  {\bf curl \,  curl}  ({\bf  e}_x \Psi)
 = {c \lambda \over \omega }
 e^{  i (  \nu  \phi + k_x x) }
\left(i {k_x \over \lambda} Z_{\nu }^{\prime} {\bf e}_r -
{k_x \nu  \over \lambda^2 r} Z_{\nu } {\bf e}_{\phi} +
Z_{\nu } {\bf e}_x \right)
\label{D034}
\end{eqnarray}
We note that $  {\bf curl} {\bf L} =0$.
{\bf L} is an expansion in eigenfunctions of the longitudinal 
electric field from the 
point source and {\bf M}  and {\bf N}  are expansions  in eigenfunctions 
of the transverse fields.

The orthogonality properties of
 functions (\ref{D031}-\ref{D034}) are 
 \begin{equation}
\int d {\bf r} 
\left(\begin{array}{cc}
{\bf L} ^{\ast} (r, \lambda ^{\prime} , k_x ^{\prime} , \nu ^{\prime} )\\
{\bf M} ^{\ast} (r, \lambda ^{\prime} , k_x ^{\prime} , \nu ^{\prime} )\\
{\bf N} ^{\ast} (r, \lambda ^{\prime} , k_x ^{\prime} , \nu ^{\prime} )
\end{array} \right) 
\left(\begin{array}{cc}
{\bf L} (r, \lambda , k_x , \nu ) \\
{\bf M} (r, \lambda , k_x  , \nu ) \\
{\bf N} (r, \lambda , k_x  , \nu )
\end{array} \right) =
 (2 \pi)^2(1 + \delta_{\nu ,0})  {\delta(\lambda -\lambda ^{\prime} )  
\over \lambda }\delta
(k_x - k_x ^{\prime} )\delta_{\nu  -\nu ^{\prime} }
\label{D501}
\end{equation}

For the particle moving along trajectory ${\bf r_0}(t) $ the 
eigenfunction 
expansion of the scalar Green's function for the 
scalar wave equation
\begin{equation}
{\bf curl \, curl\,} \Psi - {\omega^2 \over c^2} \Psi = - {4 \pi }
\int d t \, e^{ i \omega t } \, 
 {\delta (r- r_0 (t) ) \over r} \delta(x-x_0(t) )  \delta(\phi-  \phi _0(t))
\label{D0301}
\end{equation}
is 
\begin{equation}
G(r, r_0, \lambda, \nu  , k_x)=
i \pi 
  \delta ( \omega -\nu  \Omega-  k_x v_x)
\left\{\begin{array}{cc}
J_{\nu } (\lambda r )
H_{\nu }^{(1)} (\lambda r_0) & \mbox{if $ r \leq r_0 $}\\
J_{\nu } (\lambda r_0) H_{\nu }^{(1)} (\lambda  r_0) & \mbox{if $ r \geq r_0 $}
\end{array} \right.
\label{D314}
\end{equation}
This particular choice of Bessel and Hankel functions insures that the
solution is regular at zero and corresponds to the outgoing wave
at infinity.

The corresponding dyadic Green's functions for the 
vector wave equation (\ref{D729}) can be found using the
relations
\begin{eqnarray}
&&
{\cal{G} } _B ({\bf r}, {\bf r}_0) = 
{\bf curl} {\cal{I}}  G ({\bf r}, {\bf r}_0)
\mbox{} \label{D60} \\ \mbox{}
&&
{\cal{ G} } _E ({\bf r}, {\bf r}_0) =
\left({\cal{I}} + {\omega^2 \over c^2} \nabla \nabla \right)
 G ({\bf r}, {\bf r}_0)
 \label{D61a}
\end{eqnarray}
where  ${\cal{I}} $  is a unity matrix,
$ G ({\bf r}, {\bf r}_0)$ is the scalar Green's function
and ${\cal{G} } _B ({\bf r}, {\bf r}_0) $ and $ {\cal{ G} } _E ({\bf r}, {\bf r}_0)$
are magnetic and electric dyadic Green's functions.

 The  eigenfunction  expansion    of the electric  dyadic Green's function
is
\begin{eqnarray}
&&
 {\cal{G} } _E (r, r_0, \lambda, \nu  , k_x)= i \pi (2 - \delta_{\nu  , 0}) 
 \delta\left(\omega- k_{\phi} v_{\phi} - k_x u_d \right)
\mbox{} \nonumber \\ \mbox{}
&& \hskip .2 truein 
\left(
\phantom{{{{{a\over b}\over{a\over b}}}\over{{{a\over b}\over{a\over b}}}}}
{\bf L}^{\ast} ( \lambda r) \otimes {\bf L} ( \lambda r_0)+
{\bf M} ^{\ast} ( \lambda r) \otimes  {\bf M} ( \lambda r_0)+
{\bf N} ^{\ast} ( \lambda r)  \otimes {\bf N} ( \lambda r_0) \right)
\label{D0303a}
\end{eqnarray}
which, using relations (\ref{D61a}) gives
\begin{eqnarray}
&& 
{\cal{G} }_E (r, r_0, \lambda, \nu  , k_x)= i \pi (2 - \delta_{\nu ,0})
 \left(
\phantom{{{{{a\over b}\over{a\over b}}}\over{{{a\over b}\over{a\over b}}}}}
\right.
\mbox{} \nonumber \\ \mbox{}
&&
\left(H_{\nu }^{(1) \, \prime} (\lambda r) 
{\bf e}_r - {i \nu  \over  \lambda r} H_{\nu }^{(1)} (\lambda r)
{\bf e}_{\phi}-
  {i k_x \over  \lambda} H_{\nu }^{(1)} (\lambda r)  {\bf e}_x\right) \otimes
\left( J_{\nu }^{\prime} (\lambda r_0)
 {\bf e}_r + {i \nu  \over  \lambda r _0} J_{\nu }
(\lambda r_0)
{\bf e}_{\phi}+
  {i k_x \over  \lambda} J_{\nu }(\lambda r_0)  {\bf e}_x\right)
\mbox{} \nonumber \\ \mbox{}
&& 
+
{c^2 \lambda^2 \over \omega^2}
 \left(- i {k_x \over \lambda} H_{\nu }^{(1) \, \prime} (\lambda r)  {\bf e}_r -
{k_x \nu  \over \lambda^2 r} H_{\nu }^{(1)} (\lambda r) 
 {\bf e}_{\phi} +
H_{\nu }^{(1)} (\lambda r)  {\bf e}_x \right)
\mbox{} \nonumber \\ \mbox{}
&&  
\otimes
\left(i {k_x \over \lambda} J_{\nu }^{\prime} (\lambda r_0) {\bf e}_r -
{k_x \nu  \over \lambda^2 r _0 } J_{\nu } (\lambda r_0){\bf e}_{\phi} +
J_{\nu } (\lambda r_0) {\bf e}_x \right) 
\mbox{} \nonumber \\ \mbox{}
&&
+
\left.
\left(- i {\nu  \over  \lambda r} H_{\nu }^{(1)} (\lambda r) {\bf e}_r -
H_{\nu }^{(1) \, \prime} (\lambda r) {\bf e}_{\phi} \right)
\otimes
\left(i {\nu  \over  \lambda r _0 } J_{\nu } (\lambda r_0)  {\bf e}_r -
J_{\nu }^{\prime} (\lambda r_0) {\bf e}_{\phi} \right)
 \phantom{{{{{a\over b}\over{a\over b}}}\over{{{a\over b}\over{a\over b}}}}}
\right)
\label{D319}
\end{eqnarray}
The first term here is 
just the eigenfunction expansion of the Green's dyadic for the
point source.

The magnetic Green's function
(the Fourier transform in $\phi,\, x$ and $t$) for the
vector wave equation
is
\begin{equation}
 {\cal{G} } _B (r, r_0, \lambda, \nu  , k_x)= i \pi (2 - \delta_{\nu  , 0}) 
\left(
{\bf N} ^{\ast} ( \lambda r) \otimes  {\bf M} ( \lambda r_0) \, +
\phantom{{{{{a\over b}\over{a\over b}}}\over{{{a\over b}\over{a\over b}}}}}
{\bf M} ^{\ast} ( \lambda r)  \otimes {\bf N} ( \lambda r_0) \right)
\label{D0303}
\end{equation}
which, using relations (\ref{D61a}) gives
\begin{eqnarray}
&& \hskip -.3  truein
{\cal{G} } _B (r, r_0, \lambda, \nu  , k_x)= i \pi (2 - \delta_{\nu ,0}) 
{c \lambda \over \omega} 
\left(\phantom{{{{{a\over b}\over{a\over b}}}\over{{{a\over b}\over{a\over b}}}}}
\right.
\mbox{} \nonumber \\ \mbox{}
&& \hskip -.3  truein
\left(- i {\nu  \over  \lambda r} H_{\nu }^{(1)} (\lambda r) {\bf e}_r -
H_{\nu }^{(1) \, \prime} (\lambda r) {\bf e}_{\phi} \right)
\otimes
\left(i {k_x \over \lambda} J_{\nu }^{\prime} (\lambda r_0) {\bf e}_r -
{k_x \nu  \over \lambda^2 r _0 } J_{\nu } (\lambda r_0){\bf e}_{\phi} +
J_{\nu } (\lambda r_0) {\bf e}_x \right) 
\mbox{} \nonumber \\ \mbox{}
&& \hskip -.3  truein
+
\left.
 \left(- i {k_x \over \lambda} H_{\nu }^{(1) \, \prime} (\lambda r)  {\bf e}_r -
{k_x \nu  \over \lambda^2 r} H_{\nu }^{(1)} (\lambda r) 
 {\bf e}_{\phi} +
H_{\nu }^{(1)} (\lambda r)  {\bf e}_x \right)
\otimes
\left(i {\nu  \over  \lambda r _0 } J_{\nu } (\lambda r_0)  {\bf e}_r -
J_{\nu }^{\prime} (\lambda r_0) {\bf e}_{\phi} \right)
 \phantom{{{{{a\over b}\over{a\over b}}}\over{{{a\over b}\over{a\over b}}}}}
\hskip -.2 truein 
\right)
\label{D321}
\end{eqnarray}

The solution to the inhomogeneous
vector wave equation (\ref{D729})
is  given by the integral with a kernel given by the 
dyadic Green's function
\begin{equation}
{\bf F} (r,  \lambda, \nu  , k_x) = \int  r^{\prime} d  r^{\prime}
{\cal{G} }
(r, r^{\prime}, \lambda, \nu  , k_x) {\bf \cdot}  {\bf Q} (r^{\prime},
 \lambda, \nu  , k_x , \omega)
\label{D0305}
\end{equation}
where ${\bf Q} (r^{\prime},
 \lambda, \nu  , k_x)$ is a Fourier transform of  $  {\bf Q}({\bf r})$.

\subsection{Synchrotron Emissivity into Cylindrical Mode}
\label{Cy}

In this subsection we show how conventional
emissivities into plane waves can be modified to 
obtain similar relations for the emissivities into
cylindrical modes. To simplify the analysis we consider in detail
only emissivity of a charge moving along a spiral trajectory.
An important factor that simplifies
our calculations and allows one to obtain the emissivity in
a simple closed form  is that for helical motion
 a particle always stays at the
same radial coordinates. If the particles were not on the ground
gyrational level, then in 
 calculating the emissivity
it would be necessary to average the gyrational motion of 
particles taking into account the  radial dependence
of the vector normal modes.

Following the conventional approach in the theory of
electromagnetic wave-particle interaction (e.g, \cite{Melrosebook1}),
we can identify the source of energy in the wave with the 
the work done by the extraneous current:
\begin{equation}
P= \int d{\bf r}d t{\bf E} ({\bf r}, t) \cdot {\bf j} ({\bf r}, t) 
\label{w1}
\end{equation}
Using the power theorem for the Fourier transform, i.e.,
\begin{equation}
\int d x f(x) g(x) = \int {d k \over 2 \pi} f(k) g(k)
\label{w11}
\end{equation}
(where $f(k)$ and $g(k)$ are Fourier transforms of $f(t)$ and $g(t)$)
we can rewrite Eq. (\ref{w1}) for the Fourier
transforms of ${\bf E} $ and $ {\bf j}$ in time, $\phi$ and $x$
\footnote{it is also possible to make also expansion of the
vector fields  ${\bf E} $ and $ {\bf j}$ in terms of the normal
modes ${\bf L}, \, {\bf N},\, {\bf M}$ and use later the dyadic
Green's function ${\cal{G}}(\lambda, \nu, k_x, \omega)$ }
\begin{equation}
P= \sum_{\nu} \int r d r {d \omega d k_x \over (2 \pi)^3 }
 {\bf E} (r, \nu,k_x, \omega) \cdot {\bf j} ^{\ast}
(r, \nu,k_x, \omega)
\label{w13}
\end{equation}
Electric field in Eq. (\ref{w13}) may be expressed in terms of the dyadic
Green's function and the current (see Eq. (\ref{D0305})).
For a particular case of a  particle executing  helical motion with  velocity 
${\bf v} =\{0, v_\phi, v_x \}$, the Fourier transform of the current 
 density 
$ {\bf j} ({\bf r}) =q {\bf v} \delta( {\bf r} -  {\bf r}_0) $
 is
\begin{equation}
{\bf j}(r,  \lambda, \nu  , k_x, \omega)=
\int d x e^{- i k_x x} \int d t  e^{ i \omega t }  \int d \phi
e^{- i \nu  \phi} {\bf j} ({\bf r}) =
q  {\bf v}  \delta(\omega -  \nu  \Omega - k_x v_x) {\delta(r-r_0) \over r },
\label{D0307}
\end{equation}
so that the electric field is  given by ($\nu \neq 0$)
\begin{equation}
{\bf E} (r,  \lambda, \nu  , k_x, \omega) = { 2 \pi q \omega 
\over c}  \sum_i {\bf N_i} ^{\ast}
(\lambda r )
({\bf N_i} (\lambda r_0) \cdot {\bf v} ) \delta(\omega -  \nu  \Omega - k_x v_x)
\label{D030711}
\end{equation}
where ${\bf N_i} = {\bf L},\,{\bf N},\,{\bf M}$
Inserting Eq. (\ref{D030711}) in  Eq. (\ref{w13})
we find the radiated energy
\begin{equation}
P= {  2 \pi  q^2 \omega \over c}
  \sum_i \left| {\bf N_i}(\lambda r_0) \cdot {\bf v} \right|^2
\label{D030712}
\end{equation}
which gives the cyclotron emissivity
\begin{equation}
\eta(\omega , k_x) =  q^2 \omega  \sum_{\nu}
 \left(  J_{\nu } ^{\prime 2}(\lambda r_0 ) \ v_{\phi}^2 +
\left({k_x  c \over \lambda  } - 
{\omega  v_ x \over  \lambda c }
\right)^2
J_{\nu } ^2 (\lambda r_0 ) \right)  \delta(\omega -  \nu  \Omega - k_x v_x)
\label{D03010}
\end{equation}
 which is exactly 
the cyclotron emissivity per unit interval $ d \phi \,  d \omega \, d k_x $.

\subsection{Alternative Derivation of Cyclotron Emissivity}
\label{Alt}

We can use an alternative method to derive the cyclotron emissivity into 
cylindrical modes by  finding the electromagnetic fields produced by the
given charge distribution and integrate the Pointing vector over the cylindrical
surface.
Using Eqns (\ref{D319}),  (\ref{D0305}) and 
  (\ref{D0307}) we find the transverse component of the electric field:
\begin{eqnarray}
&&
{\bf E} _{\perp}=   \pi q (2 - \delta_{\nu  , 0}) {\omega \over c}
 \delta(\omega -  \nu  \Omega - k_x v_x)
\left(
- J_{\nu } ^{\prime} (\lambda r_0) v_{\phi}   
\left(- i {\nu  \over  \lambda r} H_{\nu }^{(1)} (\lambda r) {\bf e}_r -
H_{\nu }^{(1) \, \prime} (\lambda r) {\bf e}_{\phi} \right)
\right.
\mbox{} \nonumber \\ \mbox{}
&&
\left.
- {c^2 \lambda^2 \over \omega^2}
 \left({ k_x  \nu  \Omega \over  \lambda ^2}  - v_x \right) 
J_{\nu }(\lambda r_0)
\left(-  i {k_x \over \lambda} H_{\nu }^{(1) \, \prime} (\lambda r)  {\bf e}_r -
{k_x \nu  \over \lambda^2 r} H_{\nu }^{(1)} (\lambda r)
 {\bf e}_{\phi} +
H_{\nu }^{(1)} (\lambda r)  {\bf e}_x \right) \right)
\label{D0308}
\end{eqnarray}

In the wave zone, $ \lambda r \gg 1$, and for $ \nu  \gg 1$
 this can be simplified
\begin{eqnarray}
&&
{\bf E}  _{\perp} 
=i q \pi  \sqrt{{2 \over \pi \lambda  r }}  {\omega \over c}
e^{ i  \lambda  r }  
\left(
J_{\nu } ^{\prime} (\lambda r_0) v_{\phi} 
{\bf e}_{\phi}
 - {c^2 \lambda^2 \over \omega^2}
 \left(- { k_x  \nu  \Omega \over  \lambda ^2}   + v_x \right)
J_{\nu }(\lambda r_0) 
\left(i {k_x \over \lambda}  {\bf e}_r + {\bf e}_x \right)
\right) 
\mbox{} \nonumber \\ \mbox{}
&& \hskip 2 truein
 \delta(\omega -  \nu  \Omega - k_x v_x)
\label{D0402}
\end{eqnarray}

Similarly, we find for the  magnetic field 
\begin{eqnarray}
\hskip -.3 truein
&&
{\bf  B } = i \pi q (2 - \delta_{\nu ,0}) \lambda \,
\left(
\left(- i {\nu  \over  \lambda r} H_{\nu }^{(1)} (\lambda r) {\bf e}_r -
H_{\nu }^{(1) \, \prime} (\lambda r) {\bf e}_{\phi} \right)
\left( -
{k_x \nu  \over \lambda^2 r _0 } v_{\phi} + v_x
\right)  J_{\nu } (\lambda r_0)
\right.
\mbox{} \nonumber \\ \mbox{}
&&
\left.
 - \left(- i {k_x \over \lambda} H_{\nu }^{(1) \, \prime} (\lambda r)  {\bf e}_r -
{k_x \nu  \over \lambda^2 r} H_{\nu }^{(1)} (\lambda r) 
 {\bf e}_{\phi} +
H_{\nu }^{(1)} (\lambda r)  {\bf e}_x \right)
J_{\nu }^{\prime} (\lambda r_0) v_{\phi}
 \right) \delta(\omega -  \nu  \Omega - k_x v_x)
\label{D32a1}
\end{eqnarray}
which in the limit $  \lambda r \gg 1$ and $\nu  \gg 1$  gives
\begin{equation}
{\bf B} =i q \pi 
 \sqrt{{2 \over \pi \lambda  r }} \lambda
e^{ i  \lambda  r }  \delta(\omega -  \nu  \Omega - k_x v_x)
\left(i J_{\nu } (\lambda r_0)  \left(
 { k_x  \nu  \Omega \over  \lambda ^2} -  v_x \right)   {\bf e}_{\phi} +
 \left({k_x \over \lambda}  {\bf e}_r + {\bf e}_x \right) 
J_{\nu } ^{\prime} (\lambda r_0) v_{\phi} \right)
\label{D0403}
\end{equation}

The Pointing flux, $ {\bf S} = c  {\bf E}^{\ast}  \times {\bf B} /(4 \pi) $ for
$r \rightarrow \infty$ is then
\begin{equation}
{\bf S} = {q^2 \omega   \over r}
 \left(  J_{\nu } ^{\prime 2} v_{\phi}^2 + 
\left({k_x  c \over \lambda  } - 
{\omega v_ x \over \lambda c} 
\right)^2
J_{\nu } ^2 (\lambda r_0 ) \right)  \left(
{\bf e}_r + {k_x \over \lambda} {\bf e}_x  \right)
 \delta(\omega -  \nu  \Omega - k_x v_x) d l d \omega d k_x
\label{D0309}
\end{equation}
where $d l =  r d \phi$ is a unit arc length of a cylinder.

We are interested only in the radial component of the flux.
Equating the Pointing flux to the emissivity we find the  cyclotron emissivity
(\ref{D03010}).

\section{Waves in an Isotropic Dielectric}
\label{Wavesisotroipicdielectric}

We expect that the dielectric properties of a medium, i.e., collective effects,
 will
play an important  in defining the properties of the
wave-particle interaction.
So, 
as a next step, we consider in brief the waves in an 
 isotropic, frequency dispersive medium.
In such a medium the wave equation is 
\begin{equation}
{\bf curl \, curl\,  E } - {\partial^2 {\bf D } \over \partial t^2}  =0
\label{D11a}
\end{equation}
In an isotropic,
 dispersive medium the relation between the electric induction ${\bf D}$ and
and electric field may be written as 
\begin{equation}
{\bf D }(\omega) = \epsilon (\omega) {\bf E}(\omega) 
\label{D12}
\end{equation}
From Eqs. (\ref{D11}) and (\ref{D12}) it follows, that in a dielectric,the 
solutions of the wave equation are
\begin{eqnarray}
&&
{\bf E } ^{(lt)} (r, k_x,\nu , \omega) = E^{(lt)}  
\left({i k_x \over \hat{\lambda}^2} J_{\nu } ^{\prime} (\hat{\lambda} r ) {\bf e_r}-
{\nu k_x \over \hat{\lambda}^2 r} J_{\nu } (\hat{\lambda} r ) {\bf e_{\phi}} + J_{\nu } (\hat{\lambda} r )
 {\bf e_x} 
\right)
\label{D902}
\mbox{}  \\ \mbox{}
&&
{\bf B } ^{(lt)} (r, k_x,\nu , \omega) = {\omega  \epsilon
 \over \hat{\lambda}^2 c } E^{(lt)} 
\left({\nu J_{\nu } (\hat{\lambda} r )\over r}  {\bf e_r} + i 
J_{\nu } ^{\prime} (\hat{\lambda} r ) \, {\bf e_{\phi}} \right)
\label{D92}
\mbox{} \\ \mbox{}
&&
{\bf E } ^{(t)} (r, k_x,\nu , \omega)   =  E^{(t)}  \,
\left({i\nu \over r } J_{\nu } (\hat{\lambda} r ) {\bf e_r} - J_{\nu }^{\prime} {\bf e_{\phi}} 
\right)
\label{D903}
\mbox{}  \\ \mbox{}
&&
{\bf B } ^{(t)} (r, k_x,\nu , \omega) =  E^{(t)} 
\left({k_x c \over \omega}   J_{\nu }  ^{\prime} (\hat{\lambda} r ) {\bf e_r} +
{i k_x\nu  \over r \omega }  J_{\nu } (\hat{\lambda} r )  {\bf e_{\phi}}
- i { c 
\hat{\lambda}^2 \over \omega}   J_{\nu } (\hat{\lambda} r ) {\bf e_x} \right)
\label{D93}
\end{eqnarray}
where $ \hat{\lambda} = \sqrt{\omega^2/c^2 \epsilon- k_x^2} $.

\subsection{Cherenkov Emission in Cylindrical coordinates}
\label{CherenkovCylindrical}

As another didactic example 
we illustrate the features of the fields in the case of 
Cherenkov type emission taking as an example scalar waves.
The solution for the vector waves can then be obtained using the
 general relations between the solutions of the
scalar and vector wave equations.

In cylindrical coordinates the equation for the eigenfunction expansion of the
scalar Green's function is given by Eq. (\ref{D314}) with $r_0=0$. The 
only remaining term is
\begin{equation}
G( (r, r_0, \lambda, \nu  =0 , k_x)= i \pi \delta(\omega - k_x v_x) 
H_{0} (\lambda r)
\label{D94}
\end{equation}
Using the argument of the $ \delta$-function we find 
\begin{equation}
\lambda = \sqrt{{\omega^2 \epsilon \over c^2} - k_x^2 } =
 k_x  \sqrt{\epsilon \beta_x^2 -1}
\label{D95}
\end{equation}
where $\beta_x=v_x/c$.
For $   \epsilon \beta_x^2 < 1$ the argument of the 
Hankel function is complex, so that the fields decay exponentially
at large distances. For $   \epsilon \beta_x^2 >  1$ the argument of the
Hankel function is real, which corresponds to outgoing waves. 
 For $  \epsilon \beta_x^2 =1$ the  Green's function (\ref{D94}) has
a discontinuity, this is the shock front corresponding to the
Cherenkov cone.

\section{Dispersion Relation in an Infinitely Strong Curved Magnetic Field.
Vlasov Approach.}
\label{DispersionInf}

The
next  problem that we will consider is a dielectric tensor
in cylindrical coordinates for plasma in the {\it infinitely
strong} circular magnetic field.
In this particular case, it is possible to relate the electric field
and the electric induction in cylindrical coordinates
through a conventional dielectric tensor and not the 
dielectric tensor operator, in other words the dielectric tensor
becomes local.

\subsection{One Dimensional Plasma in Inhomogeneous Magnetic Field}
\label{PlasInhomo}

The equations governing the 
electrodynamics  of a plasma in a strong , weakly inhomogeneous magnetic field
 are the wave equation  
\begin{equation}
{\bf curl \, curl\,  E }  ({\bf r}, t)+ {\partial ^2 {\bf E }  ({\bf r}, t)
\over 
\partial t^2 } = {4 \pi \over c } {\partial {\bf j }
 ({\bf r}, t) \over \partial t}
\label{D0}
\end{equation}
and  collisionless  Vlasov equation 
\begin{equation}
 {\partial f ({\bf r}, t, {\bf p} )  \over \partial t} +
({\bf v} \cdot {\partial \over  \partial {\bf r} } ) f ({\bf r}, t, {\bf p}) + q
\left({\bf  E}  + {1\over c} {\bf v} \times ({\bf B+ B}^0 ) \right)
{\partial  f ({\bf r}, t, {\bf p} ) \over \partial {\bf p} } =0
\label{D01}
\end{equation}
where 
 ${\bf B}^0$ is the external magnetic field, $ f ({\bf r}, t, {\bf p} )$
is a particle distribution function, $
{\bf j }
 ({\bf r}, t)$ is a current, $q$ is a charge of particles.

We 
represent the distribution function 
as a sum of the unperturbed distribution and
fluctuating part: $ f({\bf r}, t) = f^{(0)} ({\bf r}) + f^{(1)} ({\bf r}, t)$.
We assume that the 
  unperturbed distribution function is stationary, i.e., 
$ {\partial  f^{(0)} / \partial t}  =0$. A real plasma present on the
open field lines of the pulsar magnetosphere may not satisfy this
condition. It can be hydrodynamically unstable with
a growth rate much smaller than the dynamical time (which is equal to the 
pulsar period). But since in our model problem, the plasma circulates
infinitely long along the circular magnetic field, this condition is necessary
to impose. Then the  unperturbed distribution function
satisfied the equation
\begin{equation}
{\bf v} \cdot {\partial \over  \partial {\bf r} }  f^{(0)}+
\left( {e\over c} {\bf v} \times {\bf B}^0 \right)
 {\partial  f^{(0)} \over \partial {\bf p} } =0
\label{D02a}
\end{equation}
In other words, $f^{(0)}$ must be a function of the integrals
of motion. The integrals of motion for a particle in azimuthally
symmetrical static magnetic field have been considered by 
\cite{Mikhailovskybook2}. 
The relevant constants of motion are
the radial coordinate of the gyration center
\begin{equation}
r_g = r + {v _{\phi} \over \omega_B},
\label{D010ff}
\end{equation}
parallel velocity of the  gyration center $V_{\phi}$, which,
up to  terms linear  in the small parameter
of the adiabatic approximation coincides with the parallel 
velocity of the particle $V_{\phi} = v_{\phi}$ and the quantity
$V_{\perp}$ (\cite{Mikhailovskybook2}, Eq. (15.11)), which, in turn,
up to  terms linear  in the same small parameter
 coincides with the gyrational
part of the particle velocity $V_{\perp} = v_ {\perp}$ (see 
below). 
The most general form of the distribution function is then
\begin{equation}
f^{(0)}= F(V_{\perp},V_{\phi},r_g)
\label{D02aa}
\end{equation}
where $F$ is an arbitrary function. We also note, that
since the distribution function is independent of 
$x$ and $\phi$ by our choice and cannot  depend
on the gyrational phase of the particle, the first term in
Eq. (\ref{D02a}) is zero.

Following \cite{Mikhailovskybook2}, we find, that
the inhomogeneity of the external magnetic field results in an 
equilibrium electric current (diamagnetic current) given by
(\cite{Mikhailovskybook2}) 
\begin{equation}
j_x = {c \over B} \left( {\partial P_{\perp}  \over  \partial  r} +
{  P_{\perp} -  P_{\parallel} \over r} \right)
\label{D02a1}
\end{equation}
where $  P_{\perp}$ and $  P_{\parallel}$ are transverse and 
parallel plasma pressures, defined by 
\begin{eqnarray}
&& 
P_{\perp} = \sum   \int v_{\perp} p_{\phi}f^{(0)} d {\bf v}
\mbox{} \nonumber \\ \mbox{}
&& P_{\parallel} =\sum  \int v_{\phi} p_{\phi} f^{(0)} d {\bf v}
\label{D02a2}
\end{eqnarray}
Here $ v_{\perp}^2 = v_r^2 + (v_x - u_d)^2$, $u_d$ is a drift
velocity.

Using Maxwell  equations and  Eq. (\ref{D02a1}) the condition of the 
transverse equilibrium becomes (\cite{Mikhailovskybook2}) 
\begin{equation}
{\partial  \over \partial  r}  \left(
P_{\perp} + { B^2 \over 8 \pi} \right) +
 { B^2 \over  4 \pi} + {  P_{\perp} -  P_{\parallel} \over r}  =0
\label{D02a3}
\end{equation}

In the case $P_{\perp}=0$ the diamagnetic current
(\ref{D02a1}) may be written as
\begin{equation}
j_x =q n u_d, \hskip .5 truein u_d = {\gamma_{\phi} v_{\phi}^2 \over 
\omega_B r } 
\label{D02a7}
\end{equation}
where $u_d$ is the drift velocity directed along the binormal to the
magnetic field. 
We note here,  that there is an alternative way of deriving 
Eq. (\ref{D02a7}) from the single particle equation of motion
using  averaging over  fast rotations around magnetic field
(e.g \cite{}). After averaging over fast rotation the
resulting equation of motion for the location of the gyration
center contains inertial forces due to the inhomogeneity of the
magnetic  field. These inertial forces result in a drift motion
which produces diamagnetic current.

Returning to the Vlasov equation, we  
make a Fourier transform in time,  $\phi$ and $x$ 
 ($\propto \exp \{ i( \nu \phi + k_x x-\omega t)$) and find an equation for 
$f^{(1)}( r, m, k_x, \omega) $:
\begin{equation}
i( -\omega + { v_{\phi} \nu \over r } + k_x  v_x
  + v_r  { \partial  \over \partial r} )
f^{(1)}(r, m, k_x,  \omega, {\bf p} ) +
q
\left( {\bf  E} ( r, m, k_x,  \omega) + {1\over c} {\bf v} \times {\bf B} 
( r, m, k_x,  \omega) \right)
{\partial  f ^{(0)} \over \partial {\bf p} } =0
\label{D04}
\end{equation}

\subsection{Infinitely Strong Magnetic Field}
\label{Inft}

Equation (\ref{D04}) is a {\it differential} equation for $f^{(1)}(r, m, k_x,  \omega, {\bf p} ) $.
This reflects the fact, that in cylindrical coordinates
(any curvilinear coordinates) the relation between electric
field and electric induction involves dielectric tensor
{\it operator}\footnote{Mathematically,
the difference is that dielectric tensor  acts in a space tangent to the vector
field at some point, while dielectric tensor  operator acts on a vector
field itself.}.
The case of infinitely strong magnetic field allows a considerable
simplification: in this case the operator relation is reduced to algebraic,
i.e. , to the conventional dielectric tensor.

In the case of infinitely strong magnetic field the velocities
across the field are zero: $v _x, v_r=0$ and Eq. (\ref{D04})
can be solved for  $ f^{(1)}(r, m, k_x,  \omega)$:
\begin{equation}
f^{(1)}(r, m, k_x,  \omega, {\bf p} )= 
{i q E_{\phi} \over c (\omega - \Omega  \nu) } 
{\partial   f ^{(0)} \over \partial   p_{\phi}}
\label{D041}
\end{equation}
The corresponding current density is
\begin{equation}
j_{\phi} = q \int f^{(1)}(r, m, k_x,  \omega, p_{\phi} ) v_{\phi} d p_{\phi}
= 
{q^2 i \over c} \int d p_{\phi} {v_{\phi} \over c (\omega - \Omega  \nu) } 
{\partial   f ^{(0)} \over \partial   p_{\phi}}  E_{\phi}
\label{D042}
\end{equation}
where
\begin{equation}
f^{(1)}(r, m, k_x,  \omega, p_{\phi} ) =
\int d p_x d p_r f^{(1)}(r, m, k_x,  \omega, {\bf p})
\label{D0421}
\end{equation}
is a one dimensional distribution function.

We can now introduce a dielectric tensor 
\begin{equation}
\epsilon_{ij} ( r, \nu, k_x, \omega) =
\left(
\begin{array}{ccc}
1&0&0 \\
0&1-K&0 \\
0&0& 1 
\end{array} \right)
\label{D043}
\end{equation}
where 
\begin{equation}
K= {4 \pi q^2 \over m_e } \int {d p_{\phi} \over \gamma^3}  
{f ^{(0)} \over (\omega - \Omega  \nu) ^2 } = 
{4 \pi q^2 \over \omega }  \int d p_{\phi} {v_{\phi} \over
\omega - \Omega \nu} {\partial f ^{(0)} \over \partial p_{\phi}}
\label{D044}
\end{equation}
Note, that both $j _i ( r, \nu, k_x, \omega)$ and $E_j ( r, \nu, k_x, \omega)$
are conventional Fourier transforms, so that the hermitian,
$ \epsilon_{ij}^H ( r, \nu, k_x, \omega)$, and
antihermitians, $\epsilon_{ij}^A( r, \nu, k_x, \omega)$, parts of
the dielectric tensor (\ref{D043})
satisfies the usual conditions 
\begin{eqnarray}
&& \epsilon_{ij}^H ( r, \nu, k_x, \omega) =
\epsilon_{ij}^H ( r, -\nu,-k_x,-\omega) 
\mbox{} \nonumber \\ \mbox{}
&&
\epsilon_{ij}^A( r, \nu, k_x, \omega) =
- \epsilon_{ij}^H ( r, -\nu,-k_x,-\omega)
\mbox{} \nonumber \\ \mbox{}
&&
\epsilon_{ij}( r, -\nu,-k_x,-\omega) = 
\epsilon_{ij}^{\ast}( r, -\nu,-k_x,-\omega)
\label{D04311}
\end{eqnarray}

The wave equation now reads
\begin{eqnarray}
&&
{i\nu \over r^2}  {\partial  \over \partial r} 
\left(r E_{\phi} \right) +
i k_x  {\partial \over \partial r} E_x +
\left({\nu^2 \over r^2} - \left({\omega^2 \over c^2 } -k_x^2 \right) \right)
E_r=0
\label{D400}
\mbox{} \\ \mbox{}
&&
-
{\partial \over \partial r} \left({1\over r}  {\partial \over \partial r} 
\left(r E_{\phi}  \right)  \right) +  i\nu {\partial \over \partial r} 
\left({E_r \over r } \right)  -
{k_x\nu \over r} E_x - \left({\omega^2 \over c^2 } (1-K) -k_x^2 \right) 
 E_{\phi}  =0
\label{D4001}
\mbox{} \\ \mbox{}
&&
-  {1\over r} {\partial \over \partial r}  
 \left(r {\partial \over \partial r} 
E_x  \right) +
{i k_x \over r} {\partial \over \partial r} 
\left(r E_r \right) -
\left({\omega^2 \over c^2 } - {\nu^2 \over r^2} \right) E_x =0
\label{D4002}
\end{eqnarray}

To illustrate the  solutions of Eqs. (\ref{D400} -\ref{D4002}) we   show
 how, in the WKB approximation, the solutions can be obtained for a given 
dependence $K(r)$  
in the case $k_x=0$. Then we find exact solutions for 
$k_x=0$ and constant (independent of $r$) $K$.
Unfortunately, there seems to be no simple way of solving Eqs. (\ref{D4002})
in known functions
for $k_x \neq 0$. 

\subsection{Propagation with $k_x=0$}
\label{Propk_x=0}

For $k_x=0$ the  equation for $E_x$ separates from 
the equations for $E_r$ and $E_ {\phi}$. From Eq. (\ref{D4002}) we find
\begin{eqnarray}
&&
{\bf E ^{(lt)} }({\bf r}, t) =
\exp \{i(\nu \phi -\omega t) \} J_{\nu } \left({\omega r \over c} \right) 
{\bf e_{ x}}
\label{D4003}
\mbox{} \\ \mbox{}
&&
{\bf  B  ^{(lt)} }  ({\bf r}, t) = -
\exp \{i(\nu \phi -\omega t )\} {c \over \omega} 
\left({\nu \over r} J_{\nu } \left({\omega r \over c} \right) 
{\bf e_{r}}
 + i J_{\nu } ^{\prime} \left({\omega r \over c} \right) {\bf e}_{\phi}
\right)
\label{D4004}
\end{eqnarray}

Using Eqs ({\ref{D4003}) and
  (\ref{D4004})  it is possible to show that the following relation holds:
\begin{equation}
(1-K) \, \nu 
\, E_{\phi} - i {\partial \over \partial r}\left(r  E{r} \right) =0
\label{D4005}
\end{equation}
which can be derived directly from the relation 
${\bf div \, D}=0$, where $D_i=\epsilon _{ij} E_j$ is the electric induction.

Using Eq. (\ref{D4005}) we find
\begin{equation}
{\partial^2 \over \partial r^2}  E_{r} +{1\over r} 
\left(3 + {r \over 1- K} {\partial K \over \partial r} \right)
{\partial E_{r}  \over \partial r} +
\left(\left( {\omega^{\, 2} \over c^2 } - {\nu^2 \over r^2} \right) (1-K) +
{1\over r^2} + {1 \over r (1-K) } 
{\partial K \over \partial r} \right) =0
\label{D4006a}
\end{equation}

Equation  (\ref{D4006})
can be solved using  WKB method for a given dependence
$K(r)$. If $K(r)$ is 
a constant (as a functions of $r$), then it is possible
to find  
exact solutions:
\begin{eqnarray}
&& \hskip -.3 truein
{\bf E ^{(2 )} }({\bf r}, t) =
\exp \{i(\nu \phi -\omega t) \}  \left(
{i \nu \over r} J_{\nu  \sqrt{1-K} } \left({\omega r \over c} \sqrt{1-K} \right)
{\bf e}_{r} - J_{\nu  \sqrt{1-K} }^{\prime} 
\left({\omega r \over c} \sqrt{1-K} \right) 
{\bf e}_ { \phi} \right)
\label{D4006}
\mbox{} \\ \mbox{}
&& \hskip -.3 truein
{\bf  B  ^{(t)} }  ({\bf r}, t) =
 \exp \{i(\nu \phi -\omega t )\} {i c \over \omega } 
\left({\omega^{ \, 2} \over c^2 } (1-K) +
{K \nu^2 \over r^2} \right) 
 J_{\nu  \sqrt{1-K} } \left({\omega r \over c} \sqrt{1-K} \right)
{\bf e}_{x} 
\label{D4007}
\end{eqnarray}

\subsection{Curvature Emission in Vacuum and in the Isotropic Medium}
\label{CurvatureEmission}

{ 
In this section we'll show how the conventional emissivities
for curvature emission  can be obtained in this approach.
We will calculate the single particle
emissivity 
as the energy dissipated by the optically active 
medium. This  can be done by relating in
 Eq. (\ref{w13}) 
the local current ${\bf j} ( r, \nu, k_x, \omega)$ and the
electric field  through  the antihermitian part of
the dielectric tensor (\ref{D043}):
\begin{equation}
 j _i ( r, \nu, k_x, \omega) =
- i { \omega  \over 4 \pi} \epsilon_{ij}^A E_j ( r, \nu, k_x, \omega)
{\delta( r - r_0) \over r}
\label{w191}
\end{equation}
To procede further we
define the wave amplitute, $E_{ {\bf N_i}}$, as
\begin{equation}
{\bf E}  ( r, \nu, k_x, \omega) = 2 \pi \sum_i 
  E_{ {\bf N_i}} {\bf N_i}  ( r, \nu, k_x, \lambda)
\delta(\lambda -\sqrt{ \omega^2/c^2 -k_x^2} )
\label{w192}
\end{equation}
Then the 
energy  density in
 the wave $W$ is ($V= \pi l^2 H$)
\begin{eqnarray}
&&
W = \lim_{T\rightarrow \infty}  \lim_{l\rightarrow \infty} 
\lim_{H\rightarrow \infty} {1\over T} \int _{-T/2} ^{T/2} dt 
{1\over H}
\int_{-H/2}^{H/2} d x \int_0 ^{2 \phi} d \phi 
{1\over \pi l^2}
\int_0 ^{l} r dr
 { |{\bf E} ({\bf r},t) |^2
\over 8 \pi } =
\mbox{} \nonumber \\ \mbox{}
&&
 \lim_{l\rightarrow \infty}
 \int { r dr \over \pi l^2} {1\over (2 \pi)^3}
\sum_{\nu} \int d k_x d \omega  {   |{\bf E} ( r, \nu, k_x, \omega)|^2
\over  8 \pi } = 
\mbox{} \nonumber \\ \mbox{} 
&&
\lim_{l\rightarrow \infty}
 \sum_i
 {1\over (2 \pi)^3}
\sum_{\nu} \int d k_x d \omega  { \pi | E_{ {\bf N_i}} |^2 \over 2}
\int {r dr  \over \pi l^2} | {\bf N_i}|^2  [
\delta(\lambda -\sqrt{ \omega^2/c^2 -k_x^2} )]^2
\label{w818}
\end{eqnarray}

Using the relations
\begin{eqnarray}
&&
\lim_{l\rightarrow \infty}
\int  r dr J^2 (\lambda r) = 
\lim_{l\rightarrow \infty} {l^2\over 2} \left[
J_{\nu}^{\prime\, 2}(\lambda r) + \left( 1 - {\nu^2\over \lambda^2 l^2} \right)
J_{\nu}^2(\lambda r) \right]=
{l\over \pi \lambda}
\mbox{} \nonumber \\ \mbox{}
&&
\lim_{l\rightarrow \infty} {1\over l}
[\delta(\lambda -\sqrt{ \omega^2/c^2 -k_x^2} )]^2 
 = {1\over 2 \pi} \delta(\lambda -\sqrt{ \omega^2/c^2 -k_x^2})
\label{w819}
\end{eqnarray}

We find
\begin{equation}
 W=
 {1\over (2 \pi)^3}
\sum_{\nu} \int d k_x d \omega  { | E_{ {\bf N_i}} |^2 \over 4 \pi^2 }
\delta(\lambda -\sqrt{ \omega^2/c^2 -k_x^2})
=
 {1\over (2 \pi)^3}
\sum_{\nu} \int d k_x d \omega W_{ {\bf N_i} }
\label{w193}
\end{equation}
where we identified an energy in a vector mode  ${\bf N_i}$ as
\begin{equation}
W_i = { | E_{ {\bf N_i}} |^2 \over 4 \pi^2} 
\delta(\lambda -\sqrt{ \omega^2/c^2 -k_x^2})
\label{w194}
\end{equation}
 Defining the energy gain of the wave as $- \Gamma_i W_i$
we find using Eq. (\ref{w13})
\begin{equation}
- \sum_i {1\over (2 \pi)^3}
\sum_{\nu} \int d k_x d \omega \Gamma_i W_i = - i {\omega \over 4 \pi}
{1\over (2 \pi)^3} \sum_{\nu} \int d k_x d \omega  | E_{ {\bf N_i}} |^2
\int r dr {\bf N_i} ^{\ast} (\lambda r)\cdot {\bf \epsilon} ^A \cdot {\bf N_i}
(\lambda r)
 {\delta( r - r_0) \over r}
\label{w195}
\end{equation}
so that
\begin{equation}
\Gamma_i = {\omega \over \pi} {\bf N_i} ^{\ast} (\lambda r_0)\cdot 
{\bf \epsilon} ^A \cdot {\bf N_i}
(\lambda r_0) 
\label{w196}
\end{equation}
This expression provides a growth rate for a vector cylindrical
mode ${\bf N_i} = {\bf N}, {\bf M}$ due to a current circulating at a
radius $r_0$.

To find the single particle emissivity we assume
that the growth of the wave is due to the resonant
interaction with plasma particles which emit and absorb
 wave quanta.  We then can identify the 
single particle probability of emission $w (p_{\phi}, \nu, k_x, \lambda)$
as
\begin{equation}
\Gamma  = \sum_i \int d p_{\phi} 
  w (p_{\phi},\nu, k_x,\lambda )  \hbar k_{\phi}
{\partial f ({\bf p} ) \over \partial   p_{\phi}  } 
\label{D2321}
\end{equation}
Using the
 antihermitian part of the dielectric tensor 
\begin{equation}
\epsilon^{\prime \prime}_{\phi \phi} =
- i {4 \pi^2 q^2 \over  \omega  }
\int {dp_{\phi} }  v_{\phi}
{\partial  f(p_{\phi} )  \over \partial p_{\phi} }
\delta\left(\omega-\nu \Omega \right)
\label{D2322}
\end{equation}
We find
\begin{equation}
w (p_{\phi},\nu, k_x,\lambda )= 
{4 \pi q^2 v_{\phi}^2 \over \hbar } 
\left(J^{\prime \, 2} _{\nu } (\lambda r) +
{\nu^2 k_x ^2 c^2 \over 
\lambda^2 r^2  \omega^2 } J^{2} _{\nu } (\lambda r) \right)
\delta\left(\omega-\nu \Omega \right)
\label{D2323}
\end{equation}
and emissivity $\eta (\nu, k_x,\omega)$
\begin{equation}
\eta(\omega, k_x) = 
 = \sum_{\nu } 
{\omega^3 q^2 r_0^2  \over  \nu^2 }
\left(J^{\prime \, 2} _{\nu } ( \lambda r_0) +
{\nu^2 k_x^2 c^2 \over \lambda^2  r_0^2 \omega^2 }
J^{2} _{\nu } (\lambda r  ) \right)
\delta\left(\omega-\nu \Omega \right)
\label{D2324}
\end{equation}
This is emissivity per unit range $dk_ x d \omega d \phi$.

Effectively, we calculated the induced energy loss of a particle 
due to the interaction with a wave and then, using Einstein relations
between the induced and spontaneous emission, we find the 
spontaneous emissivity.

Introducing  notations $ k_x = \omega /c \sin \theta$, 
$ d k_x = \omega /c \cos  \theta d  \theta$,
$\lambda = \omega /c \cos \theta$ and $\beta_{\phi}= \Omega r/ c $ the
expression for the curvature emissivity may be reduced to
\footnote{Since angle $\theta$ is measured from the plane $k_x=0$ the
unit solid angle is $d {\bf \Omega} = d \phi \cos \theta d \theta$}
\begin{equation}
\eta(\omega) ={\omega^2 q^2 v_{\phi}^2 \over 2 \pi c }
\left(J^{\prime \, 2} _{\nu } (\nu \beta_{\phi} \cos \theta ) +
{\tan^2 \theta \over
 \beta_{\phi} ^2 }
J^{2} _{\nu } (\nu \beta_{\phi} \cos \theta ) \right)
\delta\left(\omega-\nu \beta_{\phi} c/ r \right)
\label{D2325}
\end{equation}
This is  emissivity per unit solid angle $ d {\bf \Omega}$.
This is exactly, the expression for the single particle curvature emissivity
in vacuum
(\cite{Melrosebook2}, Eq. (13.62)).
We stress, that Eq. (\ref{D2325}) was obtained in cylindrical coordinates.
It gives a single particle emissivity per unit frequency, which is, of course,
independent of the coordinate system used.

As the waves propagate outwards from the emitting region,
the approximation of an infinitely long cylinder will cease to be true.
Then, at some transition region the cylindrical waves will become spherical.
It should be possible to consider the details of such transition
using Kirchoff integrals over a remote surface where dispersive
properties of a medium are not important. Here we only note that 
the parameter $\theta$ introduced above
corresponds in a vacuum case to the polar angle in spherical coordinates.
This provides a simple formal relation  between the spherical and 
cylindrical coordinates.

}

{\rm 
In our approach, which treats the wave-particle interaction as resonant,
the impossibility of the amplified curvature emission in vacuum,
first derived  by  \cite{Blandford1975}, follows 
from the fact that in vacuum there are no subluminous waves, so that
no waves have a phase velocity that could
fall into the region where the 
derivative $ \partial f(p_{\phi})/ \partial p_{\phi}$ is positive.
}

Similarly to vacuum case, we can obtain a curvature emissivity in an isotropic  medium:
\begin{equation}
\eta(\omega) ={\sqrt{\epsilon} \omega^2 q^2 v_{\phi}^2 \over 2 \pi c }
\left(J^{\prime \, 2} _{\nu }(\nu  \sqrt{\epsilon} \beta_{\phi} \cos \theta ) +
{\tan^2 \theta \over  \epsilon 
 \beta_{\phi} ^2 }
J^{2} _{\nu } ( \nu  \sqrt{\epsilon}  \beta_{\phi} \cos \theta ) \right)
\delta\left(\omega- \nu  \beta_{\phi} c/ r \right)
\label{D2325a}
\end{equation}
(emissivity per unit solid angle $ d {\bf \Omega}$).

\section{Airy function approximation}
\label{Airyfunction}

In this section we consider the  Airy function approximation
to the curvature emissivities of 
ultrarelativistic particles  {\it in a medium}, which is a
common way of approximating the cyclotron and
curvature emissivity of ultrarelativistic particles in vacuum 
(e.g., \cite{Melrosebook1}).  We show, that  for a
particles moving with  the  velocity larger than the
speed of light in a medium which has  $\epsilon > 1$
We assume that a particle is moving through a dielectric with the 
permittivity $\epsilon > 1$
 a  qualitatively different expansion 
of the  Airy function should be used. 

In the transition region, when the argument of the Bessel functions is 
large and 
 close to their orders it is possible to use the  Airy function approximation
\begin{equation}
J_{\nu } \left(\nu  + z \nu ^{1/3} \right) \approx 
\left( {2 \over \nu } \right)^{1/3} Ai \left(-  2 ^{1/3} z \right)=
\left\{\begin{array}{cc}
\mbox{{ \large{$ {2^{2/3} \over 3  \nu ^{1/3}} $ }} }\left(
J_{1/3} \left({ 2^{3/2} \over 3} z^{3/2}  \right)+
J_{-1/3} \left({ 2^{3/2} \over 3} z^{3/2}  \right) \right)
& \mbox{if $ z > 0$}\\ 
\mbox{{\large{$
{2^{2/3} \sqrt{z}  \over \sqrt{3} \pi \nu ^{1/3}} 
$ }} }
K _{1/3} \left({ 2^{3/2} \over 3} |z|^{3/2}  \right)
& \mbox{if $ z <  0$}
\end{array} \right.
\label{D2329}
\end{equation}
where $ Ai \left(x\right) $ is Airy function. 

{\it In vacuum}, for a resonant wave-particle interaction
 the argument of the Bessel functions is
always smaller than the order: 
$ \lambda r = \nu  \beta_{\phi} \cos \theta < \nu $. 
In a medium, the 
 argument of the Bessel functions can become large than the 
order. 
In a dielectric, the  argument of the Bessel functions is 
 $ \hat{\lambda}  r = \sqrt{\omega^2 \epsilon / c^2 -k_x^2} r$. 
If we introduce a notation $k_x = \omega \sqrt{\epsilon} / c \sin \theta$
(this is a definition of angle $\theta$, the condition of regularity
at infinity, $ \hat{\lambda}^2 >0 $ 
insures that $\sin \theta< 1$), then we have
 $ \hat{\lambda}  r = 
 \nu  \sqrt{\epsilon} \beta_{\phi} \cos \theta$, which can be
larger than $\nu $
 if $  \beta_{\phi} > 1/\sqrt{\epsilon}$ - for the superluminal
motion of a  particle.
Using these notations we find 
\begin{equation}
z= (\sqrt{\epsilon } \beta_{\phi} \cos \theta -1) \nu ^{2/3} \approx
\left(\delta -{1\over 2 \gamma^2}  -{\theta^2 \over 2} \right) \nu ^{2/3}
\label{D2330}
\end{equation}
for $ \epsilon =1 + 2 \delta$, $\delta \ll 1$, $\gamma \gg 1 $ and
$ \theta\ll 1$.
It is clear from Eq. (\ref{D2330}), that in vacuum $z< 0$ and 
in a medium $z$ becomes positive for superluminal particles.

Following the discussion of Section \ref{WKBSolution}, we can 
identify the  light cylinder radius for the mode $\nu$ as
$r_{\nu} = \nu / \hat{\lambda}$. We can 
argue, that
in the case $ z > 0$ the resonant interaction of a particle with 
a wave occurs {\it outside } the light cylinder
\begin{equation}
r_0 = r_{\nu} \left(1 + {z \over \nu^{2/3} } \right)
\label{D2331a}
\end{equation}

A transition through point $z=0$ ("light cylinder" or "classical reflection point" )
 is nontrivial. It resembles phase 
transition (\cite{Schwinger}) in a sense that correlation length for 
thermal or quantum fluctuations is very large near the 
transition. The physical conditions beyond and above
the transition  point are essentially different.

For superluminal motion the collective effects of the medium play 
important role. In this case,  the corresponding emission process   
can be called 
Cherenkov-curvature emission, stressing the fact that both 
inhomogeneous magnetic field and a medium are important for the emission.
In the vacuum limit, $ n \rightarrow 1$, 
Cherenkov-curvature emission emission reduces to the conventional 
curvature emission. Conventional Cherenkov 
radiation may be obtained in the limit $ r \rightarrow \infty$ 
after integration over $\nu $
(see  \cite{Schwinger} for the corresponding transition for the 
cyclotron-Cherenkov radiation).

The emissivity for the Cherenkov-curvature process is 
\begin{equation}
\eta(\omega) =  { 2^{2/3} \sqrt{\epsilon }\omega^2 q^2 v_{\phi}^2
\over 2 \pi c \nu ^{2/3}  } 
\left({\tan ^2 \theta  \over  \epsilon \beta_{\phi}^2}
Ai ^2 (- 2 ^{1/3} z )  +
\left({2 \over \nu } \right) ^{2/3}  Ai^{\prime \, 2} (- 2 ^{1/3} z ) 
\right) \delta\left(\omega- \nu  \Omega \right)
\label{D2331b}
\end{equation}

For $z < 0 $ this reduces to the conventional representation
for synchrotron emission in terms of MacDonald functions $K_{1/3}$.
For  $z > 0$ Eq. (\ref{D2331b}) gives 
\begin{equation}
\eta(\omega) =  { 2^{1/3} \sqrt{\epsilon }\omega^2 q^2 v_{\phi}^2 
\over 9 \pi c \nu ^{2/3}}
\left({\tan ^2 \theta  \over  \epsilon \beta_{\phi}^2}
\left(J_{1/3} (\xi) + J_{-1/3} (\xi) \right)^2 +
{2^ {2/3} z^2   \over    \nu ^ {2/3}} 
\left(J_{2/3} (\xi) - J_{-2/3} (\xi) \right)^2
\right)
\delta\left(\omega- \nu  \Omega \right)
\label{D2332}
\end{equation}
where we used 
\begin{eqnarray}
&&
 J^{\prime}_{\nu } \left(\nu  + z \nu ^{1/3} \right) \approx
\left( {2 \over \nu } \right)^{2/3} Ai ^{\prime} 
 \left(-  2 ^{1/3} z \right)
\mbox{} \nonumber \\ \mbox{}
&&
Ai ^{\prime}(-z) = -{z\over 3} \left(J_{-2/3}  \left(-  2 ^{1/3} z \right)
 - J_{2/3}  \left(-  2 ^{1/3} z \right)  \right)
\label{D2331}
\end{eqnarray}

Emissivity (\ref{D2332}) may be simplified in the case $ \xi \gg 1$.
Then we can use 
the asymptotic expansion for Airy functions
\begin{eqnarray}
&&
 Ai(- z) \approx {1\over \sqrt{\pi} z^{1/4} } \sin \left(
{2\over 3} z^{3/2} + {\pi \over 4} \right) 
\mbox{} \nonumber \\ \mbox{}
&&
Ai ^{\prime}(-z)  \approx {z^{1/4} \over \sqrt{\pi}} \cos \left(
{2\over 3} z^{3/2} + {\pi \over 4} \right)
\label{2336}
\end{eqnarray}
 to find
\begin{equation}
  \eta(\omega) = {\sqrt{2} \sqrt{\epsilon }\omega^2 q^2 \beta _{\phi}^2 
\sqrt{z} \over 
\pi^2 c \nu ^{4/3}  } \left(
\cos^2 (\xi +\pi/4) +
{\sin ^2 (\xi +\pi/4)  \nu  ^{2/3} 
\tan^2 \theta \over 2 z \epsilon \beta _{\phi}^2}
\right)
\delta\left(\omega- \nu  \Omega \right)
\label{D2333}
\end{equation}
where $ \xi = (2 z)^{3/2}/3 $.

For $\delta \gg 1/\gamma^2, \,$ and  $\delta \gg \theta^2 $ the  condition
$ z \gg 1$ implies $ \delta \nu ^{2/3} \gg 1$. Then Eq. (\ref{D2333}) 
 can be further simplified:
\begin{equation}
  \eta(\omega) \approx 
 {\sqrt{2} \sqrt{\epsilon } \sqrt{\delta} 
\omega^2 q^2 \beta _{\phi}^2   \over
\pi^2 c \nu } \cos^2 \left(
 {2 ^{3/2} \nu  \over 3} \delta ^{3/2} +{\pi \over 4} \right)
\delta\left(\omega- \nu  \Omega \right)
\label{D2334}
\end{equation}

In this context we note, that the total spectral  power
for the curvature emission in a medium, viz
\begin{equation}
\eta (\omega) = 
{q^2 \omega \over n^2 r } \left(
2 n^2 \beta _{\phi}^2 J^{\prime} _{2 \nu}
(2 \nu n  \beta _{\phi}) -
(1- n^2  \beta _{\phi}^2) \int _0^{2 \nu n  \beta _{\phi}} 
dx J _{2 \nu} (x) \right)
\label{D2335}
\end{equation}
for the case of superluminal motion, $n \beta _{\phi} > 1$, can be reduced
to the explicitly Cherenkov-type emission form:
\begin{equation}
\eta (\omega) =
{q^2 \omega \beta _{\phi} \over 4 \pi c} 
\left(1 - {1\over n^2 \beta _{\phi}^2}\right)
\Lambda(z)
\label{D2336}
\end{equation}
with $\Lambda(z)$ of the order of unity for $ z \geq 1$
(\cite{Schwinger}).

\section{Resonant 
Electromagnetic Waves in the  Asymptotic Regime $z \gg 1$}
\label{WavesAsymptotic}

We wish to simplify the relations  (\ref{D902}-\ref{D93}) in the 
limit $z \gg 1 $.   
 Instead of standing waves, described by Bessel $J$ functions, we consider
propagating waves,  described by Hankel $H^{(1)}$ and 
 $H^{(2)}$ functions. For the outgoing
waves, corresponding  to $H^{(1)}$, we find in the limit $z \gg 1$:
\begin{eqnarray}
&&
H^{(1)}  \approx - {2^{1/4} \over \sqrt{\pi} \nu ^{1/3} z^{1/4} }
e^{i {2^{3/2} \over 3} z^{3/2} + i {\pi \over 4} }
\mbox{} \nonumber \\ \mbox{}
&&
H^{(1)\, \prime}  \approx - i  {2^{3/4} z^{1/4}  \over \sqrt{\pi} \nu ^{2/3}}
e^{i {2^{3/2} \over 3} z^{3/2} + i {\pi \over 4} }
\label{D2343}
\end{eqnarray}
Next we define the radial wave number
\begin{equation}
k_r = - i {\partial \ln H^{(1)} \over  \partial r}
\label{D2344a}
\end{equation}
Using Eqs (\ref{D2343}) and (\ref{D2344a}) we find
\begin{equation}
k_r = {\sqrt{2 z} \hat{\lambda} \over \nu  ^{1/3} } \approx
 \sqrt{2 \delta}  \hat{\lambda}
\label{D2344}
\end{equation}

Introducing a notation $ k_{\phi} = \nu  /r$ the waves in the isotropic
dielectric  in the limit $ z \gg 1$ may be written 
\begin{eqnarray}
&&
{\bf E } ^{(lt)} ({\bf k}, \omega) = E^{(lt)}  
\left(
- { k_x k_r \over \hat{\lambda}^2 }  {\bf e_r}-
{k_{\phi} k_x \over \hat{\lambda}^ 2 } {\bf e_{\phi}} + 
 {\bf e_x} 
\right)
\label{D61}
\mbox{}  \\ \mbox{}
&&
{\bf B } ^{(lt)} ({\bf k}, \omega) = {\omega  \epsilon
 \over \hat{\lambda}^2 c } E^{(lt)} 
\left({ k_{\phi}  }  {\bf e_r} - k_r   
 \, {\bf e_{\phi}} \right)
\label{D62}
\mbox{} \\ \mbox{}
&&
{\bf E } ^{(t)} ({\bf k}, \omega)   =  E^{(t)}  \,
\left({ k_{\phi} }  {\bf e_r} -   k_r      {\bf e_{\phi}} 
\right)
\label{D63}
\mbox{}  \\ \mbox{}
&&
{\bf B } ^{(t)} ({\bf k}, \omega) =  E^{(t)} 
\left({  k_r  k_x c \over \omega}    {\bf e_r} +
{k_x k_{\phi} c  \over  \omega }    {\bf e_{\phi}}
-  { c 
\hat{\lambda}^2 \over \omega}    {\bf e_x} \right)
\label{D64}
\end{eqnarray}
with $ {\bf k}=\{k_r, k_{\phi},k_x \}$ and a phase dependence
$ \exp \{- i (\omega t - k_r r - k_{\phi} \phi - k_x x \}$.

With these notations the dispersion relation is
\begin{equation}
k_x^2 + k_r^2 +k_{\phi}^2 ={\omega^2 \epsilon \over c^2}
\label{D65}
\end{equation}

Relations  (\ref{D61}- \ref{D64}) look similar to the relations
 (\ref{D721}), though they were obtained in quite a different
manner. Expansions (\ref{D721}) are valid  for $ \lambda r - \nu \geq 
\nu$ with the radial wave vector defined by (\ref{D72}). 
 Expansions  (\ref{D61}- \ref{D64}) are valid  for $ \lambda r \approx
\nu$  and $\xi \gg 1$ (the radial wave vector in this case
is  defined by (\ref{D2344})).

The importance of these results is that in these two  limits, the
electromagnetic wave in cylindrical coordinates look like plane waves.
This is a considerable simplification.  It allows one to implement
a well developed technique of Fourier transforms in considering the 
wave propagation.
This representation of the electromagnetic fields requires (i) the presence
of a medium with the index of refraction $ n > 1$, (ii) superluminal 
motion of the resonant particle $ \delta > 1 /(2 \gamma^2) $, (iii)
 large harmonic number $ \nu  > \delta ^{-3/2}$.

Another important feature is that in this limit there is an additional
freedom in the choice of wave polarizations
in an isotropic dielectric. 
For example,
we can introduce the polarization vector corresponding to the
waves (\ref{D61}-\ref{D64}):
\begin{eqnarray}
&&
{\bf e} ^ {(t)}= {1 \over  \hat{\lambda} } \{k_{\phi} , - k_r,0\}=
 \{\sin \psi, - \cos  \psi, 0\}
\mbox{} \nonumber \\ \mbox{}
&&
{\bf e} ^ {(lt)}= 
 \{- \cos  \psi \sin \theta , - \sin \psi \sin \theta , 
\cos  \theta \}
\label{D66}
\end{eqnarray}
which satisfy relations
\begin{eqnarray}
&&
{\bf e} ^ {(t)}= {\bf e}_{\bf k} \times {\bf e}_x 
\mbox{} \nonumber \\ \mbox{}
&&
{\bf e} ^ {(lt)} = {\bf e}_{\bf k} \times {\bf e} ^ {(t)}
\label{D67}
\end{eqnarray}

The special role played by the ${\bf e}_x$ in defining the polarizations
of the normal modes comes from the fact that in the isotropic homogeneous
medium the solutions of the vector wave equations may be chosen to be
tangential to the coordinate surfaces ${\bf e}_x$. 
In the nonisotropic medium this is not true (see Sec. \ref{pl}).

Alternatively, {\it in the limit $ z \gg 1$} we can chose the
following polarizations 
\begin{eqnarray}
&&
{\bf e} ^ {(t)}=  {1 \over k_{\perp} } \{- k_x, 0, k_r \}
\mbox{} \nonumber \\ \mbox{}
&&
{\bf e} ^ {(lt)} = {1 \over k k_{\perp} } \{k_{\phi} k_r,
- k_{\perp}^2 , k_x k_{\phi} \}
\label{D68}
\end{eqnarray}
(see Fig. \ref{Polariz}).
This particular choice of polarizations  has an advantage that it may
be related to the polarizations in the straight field line geometry.

\section{Response Tensor for a One Dimensional Plasma in a Curved 
Magnetic Field}
\label{ResponceTensor}

We will calculate the response tensor of a one dimensional plasma
in cylindrical coordinates using the analogy of the forward-scattering
method (see  e.g. \cite{Melrosebook2}) adopted to cylindrical coordinates.
We represent the current as a sum of currents due to the
each single particle moving along its trajectory:
\begin{equation}
{\bf j}({\bf r} , t) =  \int d {\bf r} ^0 d {\bf p} \,
 {\bf j}_{sp} ({\bf r} , t)
f( {\bf p}, {\bf r}^0 (t) )
\label{D30}
\end{equation}
We expand the single particle current (denoted by the subscript $sp$)
\begin{equation}
{\bf j}_{sp} ({\bf r} , t)
 = q  \dot{{\bf r} } \, \delta({\bf r}  -  {\bf r}^0(t) )
\label{D31}
\end{equation}
in Fourier amplitudes in time, $k_x$ and $\phi$ and in  Hankel amplitudes
in $r$ 
\begin{eqnarray}
&&
{\bf j}_{sp}(r, m, k_x, \omega) = q
\int dt \exp \{- i \omega t\} \int dx  \exp \{i  k_x x\} \int d \phi
 \exp \{  i \nu \phi \} \, {\bf j} _{sp} ({\bf r} , t) 
\mbox{} \nonumber \\ \mbox{}
&& = \int dt
\exp \{i(\nu \phi^0(t)  +  k_x x_0(t) - \omega t ) \}  \dot{{\bf r}}
\int \xi d \xi J_{\nu }(\xi r) J_{\nu }(\xi r^0(t) )
\label{D32}
\end{eqnarray}
where we used a Hankel transform of the delta function
\footnote{
A Hankel transform is defined as 
$ F(x)=\int_0^{\infty} J_{\nu}(\xi x) \xi 
d \xi \int_0^{\infty} y dy J_{\nu}(\xi y) F(y)
$}
\begin{equation}
 \int \xi d \xi J_{\nu }(\xi r) J_{\nu }(\xi r^0) ={
\delta(r- r^0)\over r}
\label{D40}
\end{equation}
We expand the orbit of the particle in powers of wave amplitudes:
\begin{equation}
{\bf r}(t) = {\bf r}^0(t) +  {\bf r}^{(1)}(t) 
\label{D33}
\end{equation}

The first order  Fourier  transform  of the single particle current
is then
\begin{eqnarray}
&& \hskip -.2 truein
{\bf j}_{sp}^{(1)} (r, m, k_x, \omega) =
q \int dt \exp \{i(\nu \phi^0(t)  +  k_x x_0(t) - \omega t \}
\int \xi d \xi J_{\nu }(\xi r)
\mbox{} \nonumber \\ \mbox{}
&&  \hskip -.2 truein
\times
\left[\dot{ {\bf r}} ^{(1)} J_{\nu }(\xi r^0(t) ) +  {\bf r}^0(t)
\left( \left(i\nu  \phi ^{(1)} (t) + i k_x x^{(1)} (t) \right)
J_{\nu }(\xi r^0(t) ) + {\partial J_{\nu }(\xi r^0(t) ) \over \partial r^0} 
r^{(1)} (t) \right) \right]
\label{D34}
\end{eqnarray}

The orbit of a particle is found by solving the equation of motion
\begin{equation}
{d  {\bf p} \over d t} = {\bf F}^0 (t, {\bf r}, {\bf v})+
 {\bf F} ^{(1)} (t, {\bf r}, {\bf v})
\label{D35}
\end{equation}
where ${\bf F} ^0 (t, {\bf r}, {\bf v})$ is a force acting on a particle
when no wave is present and $  {\bf F} ^{(1)} (t, {\bf r}, {\bf v})$
i.e. a force acting on a particle
due to the presence of a waves.
 Expanding Eq. (\ref{D35}) in powers of waves amplitudes
we find 
\begin{eqnarray}
&&
{d  {\bf p} ^0 (t) \over d t} = {\bf F}^0 (t, {\bf r}^0, {\bf v}^0)
\label{D351}
\mbox{} \\ \mbox{}
&&
{d  {\bf p}  ^{(1)} (t) \over d t} = {\bf F}^{(1)}  (t, {\bf r}^0, {\bf v}^0)
\label{D36}
\end{eqnarray}

The force $ {\bf F}^{(1)}  (t, {\bf r}^0, {\bf v}^0)$ can be expanded in 
Fourier amplitudes 
\begin{equation}
 {\bf F}^{(1)}  (t, {\bf r}^0, {\bf v}^0) =
\sum_{\nu^{\prime}}
\int d\omega^{\prime} d k_x^{\prime} 
\exp \{-i (\omega^{\prime} t - k_x^{\prime} x^0(t) - \nu^{\prime} \phi^0(t)) 
{\bf F}^{(1)}  (\omega, m, k_x, r^0(t), t)
\label{D37}
\end{equation}

Equation (\ref{D36}) can be solved for the
first order velocity perturbation $ \dot{ {\bf r}} ^{(1)}(t)$ and
first order trajectory perturbations
$ x ^{(1)} (t),\, \phi^{(1)} (t)$ and $ r^{(1)} (t)$:
\begin{equation}
\left(\begin{array}{ll}
 \dot{{\bf r}} ^{(1)}(t)  &\\
x ^{(1)} (t)& \\
\phi^{(1)} (t) &\\
r^{(1)} (t)
\end{array} \right) =
\sum_{\nu^{\prime}}
\int d\omega^{\prime} d k_x^{\prime}
\exp \{-i (\omega^{\prime} t - k_x^{\prime} x^0(t) - \nu^{\prime} \phi^0(t))
\left(\begin{array}{ll}
 \tilde{{\bf V}} 
(\omega ^{\prime}, \nu^{\prime}, k_x ^{\prime} , r^0(t), t) & \\
\tilde{x} ^{(1)}  (\omega ^{\prime},\nu ^{\prime}, k_x ^{\prime}, r^0(t), t)
& \\
\tilde{\phi}^{(1)} (\omega ^{\prime},\nu ^{\prime} , k_x ^{\prime}, r^0(t), t)
 &\\
\tilde{r}^{(1)} (\omega ^{\prime} ,\nu ^{\prime}, k_x ^{\prime}, r^0(t), t)
\end{array} \right) 
\label{D38}
\end{equation}
where tilde denotes Fourier transforms.

The first order single particle current then becomes
\begin{eqnarray}
&&
{\bf j}_{sp}^{(1)} (r, m, k_x, \omega) =
 q \sum_{\nu^{\prime}} \int d\omega^{\prime} d k_x^{\prime} \int dt 
e^{\{i \left( ( \omega - \omega^{\prime})t -
(k_x - k_x^{\prime}) x^0(t) - (\nu - \nu^{\prime}) \phi^0(t)) \right) \}}
 \int \xi d \xi J_{\nu }(\xi r)
\mbox{} \nonumber \\ \mbox{}
&&
\left[{\bf \tilde{V} }  J_{\nu }(\xi r^0(t) ) +  {\bf r}^0(t)
\left( \left(i\nu \tilde{\phi}^{(1)} + i k_x \tilde{x} ^{(1)} \right)
J_{\nu }(\xi r^0(t) ) +
 {\partial J_{\nu }(\xi r^0(t) ) \over \partial r^0}
\tilde{r}^{(1)} \right)  \right]
\label{D39}
\end{eqnarray}

Using this single particle current we can calculate
the current (\ref{D30}).
In the general case, current (\ref{D39}) may be related to the
electric field of the perturbing wave through a generalized dielectric
tensor:
\begin{equation}
{\bf j} (r, m, k_x, \omega)  = {\cal{E}} (r, m, k_x, \omega) \cdot {\bf E}
(r, m, k_x, \omega)
\label{D81}
\end{equation}
where ${\cal{E}}$ is a dielectric tensor  operator which involves a Hankel 
transform of the external  current and  partial derivatives with respect to $r$.
This is different from Cartesian coordinates, where the electric induction
is related to the electric field through a dielectric tensor \footnote{Mathematically,
the difference is that dielectric tensor  acts in a space tangent to the vector
field at some point, while dielectric tensor  operator acts on a tensor 
filed itself.}.
To simplify  the calculations we assume that the radial dependence
of both the Hankel transform of the external  current and of the
perturbing electric field can be approximated by the plane wave form.
For the nonresonant term in the  dielectric tensor  operator this 
is justified  in the WKB limit, while for the 
resonant terms this is justified for subluminous waves in the limit
$ \delta \nu ^{2/3}  \gg 1$.

Thus, in the Hankel transform of the external  current
and in the derivatives of the perturbing electric field,
we can identify
\begin{equation}
{\partial  \over \partial r^0} \rightarrow i k_{r_0} 
,\hskip .2 truein
{\partial  \over \partial r} \rightarrow i k_r
\label{D82}
\end{equation}
In this approximation the  Hankel
transform of the external  current will reduce to the
Fourier transform in $r$ and the derivatives of the perturbing
electric field are replaced by $ i k_r$.
The dielectric tensor  operator $ {\cal{E}} $ then becomes a
conventional  tensor.

The integration of the induced current 
over $ d \phi^0  d x ^0 $ then  gives
delta-functions $\delta (\nu- \nu^{\prime}) \, \delta (k_x - k_x^{\prime}) $
that are subsequently removed by the corresponding integrations.
For $r_0 = {\rm const}$, the time and $d \omega^{\prime}$ 
integrations insure that only secular
terms are retained.
 The integration over $ d \xi$ and $ d r^0$ are removed using
relation (\ref{D40}).

\subsection{Perturbed Trajectory}
\label{PerturbedTrajectory}

The equations of motions for a particle in a circular magnetic field
when the Larmor radius is much smaller than the radius of curvature are
the following:
\begin{eqnarray}
&& 
{d p_r \over d t} = e E_r({\bf r}^0)+ {q \over c} 
\left((B_{\phi}({\bf r}^0) + B_0)
 v_x - v_{\phi} B_x ({\bf r}^0) \right)  
\label{D4110}
\mbox{} \\ \mbox{}
&&
{d p_x \over d t} = e E_x ({\bf r}^0) + {q \over c} 
\left(B_{r} ({\bf r}^0) v_{\phi} - v_{r}(B_{\phi} ({\bf r}^0) + B_0)
 \right) 
\label{D4111}
\mbox{} \\ \mbox{}
&&
{d p_{\phi} \over d t} = e E_{\phi} ({\bf r}^0) + {q \over c} 
\left(B_{x} ({\bf r}^0) v_{r} - v_{x} B_{r} ({\bf r}^0) \right) 
\label{D4112}
\end{eqnarray}

We solve Eqs (\ref{D4110}- \ref{D4112}) by expanding in powers of the wave
amplitudes.  We assume that initially the particles are in the ground gyration
state. Then
in the zeroth order we find: 
$v_r^0=0,\, {\phi}^0 =  \Omega t=  v_{\phi} t /r ,\, v_x^0 = u_d $,  with 
$ u_d ={ \gamma v_{\phi}^2 \over R_B \omega_B}$.
The first order in wave amplitudes gives
\begin{eqnarray}
&& 
{d v_r ^{(1)} \over d t} = {q \over \gamma m_e c} 
\left(E_r({\bf r}^0)-{v_{\phi} \over c} 
 B_x ({\bf r}^0)
  + {u_d \over c} B _{\phi} ({\bf r}^0)
 \right) +{\omega_B \over \gamma} v_x^{(1)} 
\label{D4116}
\mbox{} \\ \mbox{}
&&
{d v_x ^{(1)} \over d t} =  {q \over \gamma m_e c} \left(E_x ({\bf r}^0) + 
{v_{\phi} \over c} 
B_{r}  ({\bf r}^0)
  - {u_d (v_{\phi} E_{\phi}  ({\bf r}^0) + u_d E_x ({\bf r}^0)
 )\over c^2}  \right)
 + {\omega_B \over \gamma} v_r^{(1)}
\label{D4115}
\mbox{} \\ \mbox{}
&&
{d v_{\phi}  ^{(1)} \over d t} = 
 {q \over \gamma m_e c} \left( E_{\phi} ({\bf r}^0)
 - 
{u_d \over c} B _r ({\bf r}^0) -  {v_{\phi}    (v_{\phi} E_{\phi} ({\bf r}^0)
 + u_d E_x ({\bf r}^0) )\over c^2} 
\right)
\label{D4313}
\end{eqnarray}

Expanding Eqs (\ref{D4116} - \ref{D4313}) in Fourier amplitudes
we find the 
  solutions:
\begin{eqnarray}
\hskip -.2 truein 
&&
\tilde{v}_r ^{(1)} =
 { q \over m_e c \gamma ( \omega_B^2 /\gamma^2 - \Omega^{0 \, 2} )}
\left( i \Omega^0 \left( E_r + { B_{\phi} u_d - B_x v_{\phi}\over c} 
\right)  \right.
\mbox{} \nonumber \\ \mbox{}
&& \left. \hskip 1 truein +
{\omega_B \over \gamma} \left( E_x + { v_{\phi} B_r \over c} -
{ u_d  ( v_{\phi} E_{\phi} + u_d E_x )\over c^2} 
  \right) \right)
\label{D4116a}
\mbox{} \\ \mbox{}
&&
\tilde{v}_x^{(1)} =
 { q \over m_e c \gamma ( \omega_B^2 /\gamma^2 - \Omega^{0, 2})}
\left( i \Omega^0  \left( E_x + { v_{\phi} B_r \over c} -
{ u_d ( ( v_{\phi} E_{\phi} + u_d E_x )\over c^2}  \right) 
 \right.
\mbox{} \nonumber \\ \mbox{}
&& \left.  \hskip 1 truein
-
{\omega_B \over \gamma} \left( E_r + { B_{\phi} u_d -  B_x v_{\phi}\over c} 
\right)  \right)  
\label{D4117}
\mbox{} \\ \mbox{}
&&
\tilde{v}_{\phi} ^{(1)} = i { q \over m_e \Omega^0 \gamma }
\left( - \left( E_{\phi} - {u_d \over c} B _r \right) +
  { v_{\phi}    ( v_{\phi} E_{\phi} + u_d E_x )\over c^2}  \right)
\label{D4118}
\end{eqnarray}
where $ \Omega^0 = \omega -k_{\phi} v_{\phi} - k_x u_d$,
$ k_{\phi} = m/r$.

The first order variations in trajectory are 
\begin{equation}
x ^{(1)} = {v_x ^{(1)} \over i  \hat{\omega} },
\hskip .2 truein 
\phi  ^{(1)} = {v_{\phi}  ^{(1)} \over i  \hat{\omega} r },
\hskip .2 truein
r ^{(1)} = {r  ^{(1)} \over i  \hat{\omega}}
\label{D4119}
\end{equation}

In cylindrical coordinates, the relation between the electric and magnetic field
is
\begin{eqnarray}
&&
B_r= {k_x c \over \omega} E_{\phi} - {k_{\phi}  c\over \omega} E_x
\mbox{} \nonumber \\ \mbox{}
&&
B_{\phi} = - {k_x c \over \omega} E_r - i {c \partial E_x \over  \omega \partial r}
\mbox{} \nonumber \\ \mbox{}
&&
B_x =  {k_{\phi}  c\over \omega} E_r + i {c\over \omega r }
{\partial  \over \partial r} \left(r  E_{\phi} \right)
\label{D4120}
\end{eqnarray}

\subsection{Simplified Response Tensor}
\label{Simplified}

Using Eqns (\ref{D4116a} - \ref{D4120}) we can find the dielectric tensor operator
${\cal{E}}$.
In Sections \ref{ShortWaveLength} and  \ref{WKBSolution}
we discussed when the operator 
relations between electric field and electric displacement can be
simplified to algebraic relations for the nonresonant waves,
and in Section \ref{WavesAsymptotic} we found a limiting case
$ z \gg 1$
when this can be done for the resonant waves.
In these limits it is possible to change all the radial derivative
$ \partial_r \rightarrow i k_r$.
The relations between magnetic and electric field then simplify
\begin{eqnarray}
&&
B_{\phi}
\approx - {k_x c \over \omega} E_r + {k_r c \over \omega}   E_x
\mbox{} \nonumber \\ \mbox{}
&&
B_x
\approx  {k_{\phi}  c\over \omega} E_r - {k_r c \over \omega} E_{\phi}
\label{D4120a}
\end{eqnarray}

The resulting dielectric tensor is still complicated.
It can be simplified according to the following procedure:
(i) for the nonresonant parts the drift velocity is small and can be neglected,
(ii) for Cherenkov-type resonances (resonances that do not involve
$\omega_B$) we retain all the terms,
(iii) for cyclotron-type  resonances (resonances that do involve
$\omega_B$) we assume that $k_x /k_{\phi} \approx  u_d /c$, 
$ u_d /c \gg 1/\gamma$ and  expand in small drift velocity
 retaining only the first  terms in $ u_d$ and $k_x$.
Implicit in this expansion in small $u_d$ is the assumption
that $ \gamma \omega^2 \gg \omega_B^2 u_d^2/c^2$.
 In this limit, in the cyclotron-type  resonant terms involving drift velocity
we can approximate $v_{\phi} \approx c$.
The 
 dielectric tensor is then
 \begin{eqnarray}
\epsilon_{xx}&&=1-{1\over 2} \sum _{\alpha}
 \,{\omega_{p \alpha}^2\over\omega^2 }\,
\int {d p_{\phi}\over \gamma}  \left((\omega- k_{\phi} v_{\phi} )
A^+_{\alpha}
 f_{\alpha}  
 \right) -\sum _{\alpha} \,{\omega_{p \alpha}^2 }\,
 \int {d p_{\phi}\over \gamma}  {f_{\alpha} \over
\Omega^{o \,2}_{\alpha}  } { k_{\phi}^2  u _{\alpha} ^2  \over \omega^2}
\left(1 - {\omega v_{\phi} \over k_{\phi}  c^2 } \right)
\mbox{} \nonumber \\ \mbox{}
\epsilon_{rr}&&=1-{1\over 2} \sum _{\alpha}
 \,{\omega_{p \alpha}^2\over\omega^2 }\,
\int {d p_{\phi} \over \gamma}
  (\omega- k_{\phi} v_{\phi} -k_x u_{\alpha}) A^+_{\alpha}
 f_{\alpha}
\mbox{} \nonumber \\ \mbox{}
\epsilon_{{\phi} {\phi}}&&=1-\sum _{\alpha} \,{\omega_{p \alpha}^2 }\,
 \int {d p_{\phi}\over \gamma}  {f_{\alpha} \over 
\Omega^{o \,2}_{\alpha}  } \left( ( 1 -  {k_x u_{\alpha} \over  \omega }) 
\left((1 -  {k_x u_{\alpha} \over  \omega }) - {v_{\phi}^2 \over c^2}  \right) \right)
 \mbox{} \nonumber \\ \mbox{}
&&
-\sum _{\alpha} \,
{\omega_{p \alpha}^2\over \omega^2}\,
\int {d p_{\phi}\over \gamma}  f_{\alpha}   \,
{(k_x^2+ k_r^2) \, v_{\phi}^2 \, \over 
\Omega _{\alpha}^+ \Omega_{\alpha}^- }-
{1\over 2} \sum _{\alpha}
\,{\omega_{p \alpha}^2\over \omega^2}
 \int {d p_{\phi}\over \gamma} k_r u_{\alpha} 
 A^-_{\alpha} 
f_{\alpha}
\mbox{} \nonumber \\ \mbox{}
\epsilon_{xr}&&=-{i\over 2} \sum _{\alpha}
\,{\omega_{p \alpha}^2\over \omega^2}
 \int {d p_{\phi}\over \gamma} \left((\omega- k_{\phi} v_{\phi} ) A^-_{\alpha} + 
i  k_r u_d A^+ _{\alpha} \right)
 f_{\alpha}
\mbox{} \nonumber \\ \mbox{}
\epsilon_{rx}&&= {i\over 2} \sum _{\alpha}
\,{\omega_{p \alpha}^2\over \omega^2}
 \int {d p_{\phi}\over \gamma} \left(
 (\omega- k_{\phi} v_{\phi} ) A^-_{\alpha} -
 i k_r u_{\alpha}  A^+ _{\alpha} \right)
 f_{\alpha}
\mbox{} \nonumber \\ \mbox{}
\epsilon_{x\phi}&&=-{1\over 2} \sum _{\alpha}
\,{\omega_{p \alpha}^2\over \omega^2 }
 \int {d p_{\phi}\over \gamma} v_{\phi} 
 \left(  \left(k_x  - {\omega u_{\alpha} \over c^2} \right)  A^+_{\alpha} +
i k_r    A^-_{\alpha} \right )   f_{\alpha} 
\mbox{} \nonumber \\ \mbox{}
&&
-\sum _{\alpha}
\,{\omega_{p \alpha}^2\over \omega^2 }
 \int {d p_{\phi}\over \gamma} {v_{\phi} u_{\alpha} }
{ \left(k_x^2 +k_r^2\right)  \over 
\Omega _{\alpha}^+ \Omega_{\alpha}^- }  f_{\alpha} 
\mbox{} \nonumber \\ \mbox{}
&&
-  \sum _{\alpha} \,{\omega_{p \alpha}^2 }\,
 \int {d p_{\phi}\over \gamma}  {f_{\alpha} \over
\Omega^{o \,2}_{\alpha}  } {k_{\phi} u_\alpha \over \omega                                                                                                                                                                                        } \left(
( 1 -  {k_x u_{\alpha} \over  \omega }
) - {v_{\phi}^2 \over c^2 } \right)
\mbox{} \nonumber \\ \mbox{}
\epsilon_{\phi x} &&= {1\over 2} \sum _{\alpha}
\,{\omega_{p \alpha}^2\over \omega^2 }
 \int {d p_{\phi}\over \gamma} v_{\phi} 
 \left( k_x  A^+_{\alpha} -
 i k_r A^-_{\alpha} \right )  f_{\alpha} 
\mbox{} \nonumber \\ \mbox{}
&& -
 \sum _{\alpha} \,{\omega_{p \alpha}^2 }\,
 \int {d p_{\phi}\over \gamma}  {f_{\alpha} \over
\Omega^{o \,2}_{\alpha}  }  { k_{\phi} u_\alpha \over \omega}
 \left(1 -  {k_x u_{\alpha} \over  \omega } 
\right) \left( 1 - {\omega v_{\phi} \over k_{\phi}  c^2 } \right)
\mbox{} \nonumber \\ \mbox{}
\epsilon_{r\phi}&&= - {1\over 2} \sum _{\alpha}
\,{\omega_{p \alpha}^2\over \omega^2}
 \int {d p_{\phi}\over \gamma}{v_{\phi} \over c}  \left(
k_r A^+_{\alpha} - i \left(k_x c  - {\omega  u_{\alpha}  \over c} \right)  A^-_{\alpha}
\right) f_{\alpha}
\mbox{} \nonumber \\ \mbox{}
\epsilon_{\phi r} &&= - {1\over 2} \sum _{\alpha}
\,{\omega_{p \alpha}^2\over \omega^2}
 \int {d p_{\phi}\over \gamma}{v_{\phi} \over c}  \left(
k_r A^+_{\alpha} +  i  k_x  A^-_{\alpha} \right)
 f_{\alpha}
\label{epsilon}
\end{eqnarray}
Here
\begin{equation}
A^+_{\alpha}=\left({1\over \Omega^+ _{\alpha} }+{1\over \Omega^-
 _{\alpha} } \right),
\hskip .2truein
A^-_{\alpha}=\left({1\over \Omega^-_{\alpha} }-{1\over \Omega^+
_{\alpha} } \right),
\hskip .1truein
\Omega^{\pm}_{\alpha}
=\omega - k_{\phi} v_{\phi}  - k_x u_{\alpha}  \pm \omega_{B } \gamma^{-1},
\hskip .1truein
\Omega^o _{\alpha} =\omega - k_{\phi} v_{\phi}  - k_x u_{\alpha},
\label{As}
\end{equation}
 where $f_{\alpha}$ are one dimensional distribution functions of
the components ${\alpha}$.

This dielectric tensor reduces to the dielectric tensor for plasma
in straight magnetic fields for $ u_{\alpha}  =0$.
It takes a correct account of the Cherenkov-curvature emission
and gives the drift corrections to the cyclotron emission.
We also note that this dielectric tensor is nonhermitian, since 
$k_r$ is not a Killing vector.

\section{Waves in Cylindrical Coordinates in Anisotropic Dielectric}
\label{pl}

In this section we consider polarization of waves in 
anisotropic dielectric in cylindrical coordinates.
Since the dielectric properties of  a medium are determined
by the nonresonant wave-particle interaction, we assume that it can
be considered in the WKB approximation, so we can use the dielectric 
tensor (\ref{epsilon}).
The properties of a  inhomogeneous medium
are determined by the very strong, circular magnetic field.

\subsection{Infinitely Strong Magnetic Field}
\label{pl1}

In the infinitely strong magnetic field the dielectric tensor
is (\ref{D0421}). The
dispersion relation is then
\begin{equation}
\left(
(1 - K ) (1 - n_{\phi}^2) - n_r^2 -n_x^2 \right) 
(1 - n_{\phi}^2 - n_r^2-  n_x^2) =0
\label{D50}
\end{equation}
where $ n_i = k_i c /\omega$.
Eq. (\ref{D50}) has solutions
\begin{eqnarray}
&&  
n^2 =1 , \hskip 1 truein
 {\bf e}^{(t)} = {1 \over n_{\perp}} \{- n_x, 0, n_r \};
\mbox{} \nonumber \\ \mbox{}
&&  
  n_{\phi}^2 = 1- {n_{\perp}^2 \over 1- K} ,  
\mbox{} \nonumber \\ \mbox{}
&&  
{\bf e}^{(lt)} = 
{1 \over \sqrt {(1-K)^2 + K n_{\perp}^2} }
\mbox{} \nonumber \\ \mbox{}
&& \hskip .3 truein \times
\left\{
 \sqrt {(1-K) (1-K - n_{\perp}^2) } {n_r \over n_{\perp}}, - n_{\perp},
 \sqrt {(1-K) (1-K - n_{\perp}^2) } 
{n_x  \over n_{\perp}} \right\}
\label{D51}
\end{eqnarray}
where ${\bf e}^{(t)}$ and ${\bf e}^{(lt)}$ are polarization vectors
of the $t$ and $lt$ mode according to classification of  \cite{Kazbegi}.
The electric field of the $t$-wave is always perpendicular to the
magnetic field and the wave vector.

In the limit $ K \rightarrow 0$ the polarization vector for the $lt$ wave
reduces to 
\begin{equation}
{\bf e}^{(lt)} =
{1\over n_{\perp}} \{- n_r n_{\phi} , n_{\perp} , -n_x   n_{\phi} \}
\label{D511}
\end{equation}

\subsection{Finite Magnetic Field}
\label{pl2}

In the finite magnetic field the dielectric tensor can be found from Eq. (\ref{epsilon}).
For a cold plasma in the center of momentum frame the 
 dielectric tensor is
\begin{equation}
\epsilon_{ij} =
\left(
\begin{array}{ccc}
1+2 d &0&0 \\
0&1-K&0 \\
0&0& 1+ 2 d
\end{array} \right)
\label{D043a}
\end{equation}
where $ d=  \omega_p^2 /\omega_B^2$. Eq. (\ref{D53}).

Dispersion relation is then
\begin{equation}
\left(
- (1 - K ) n_{\phi}^2 + (1 + 2  d)  (1  - K  - n_r^2 -n_x^2) \right)
(1 + 2 d - n_{\phi}^2 - n_r^2 - n_x^2) =0
\label{D53}
\end{equation}
where $ d=  \omega_p^2 /\omega_B^2$. Eq. (\ref{D53})
 has solutions
\begin{eqnarray}
&&
n^2 =(1+ 2 d), \hskip 1 truein 
 {\bf e}^{(t)} = {1 \over n_{\perp}} \{- n_x, 0, n_r \}
\mbox{} \nonumber \\ \mbox{}
&&   n_{\phi}^2 = (1+ 2 d) \left(1 - {n_{\perp}^2 \over 1- K} \right),
\mbox{} \nonumber \\ \mbox{}
&& {\bf e}^{(lt)} \approx 
 {1 \over \sqrt {(1-K)^2 + K n_{\perp}^2}}
\{\sqrt {(1-K) (1-K - n_{\perp}^2) } {n_x \over n_{\perp}}
, - n_{\perp},
{n_r  \over n_{\perp}} \}
+ O(d) 
\label{D54}
\end{eqnarray}
So that the polarization vectors are the same as in the case of infinitely
strong magnetic field within factors $ \omega_p^2 / \omega_B^2$.

\subsection{Growth rate of the Cherenkov-drift instability}
\label{Growthratedrift}

Next we calculate the growth rate of the Cherenkov-drift instability
(\cite{LyutikovMachabeliBlandford1}) of a beam
propagating through a medium.
In the plane-wave approximation the  resonance condition is
\begin{equation}
{1 \over 2 \,  \gamma_{res}^2 } - \delta +
{k_r^2  \over 2 k_{\phi}^2 }  +
{1 \over 2} \left({k_x \over k_{\phi}}  - {u_d\over c } \right)^2 = \,0 ,
\label{aa11}
\end{equation}
where we used $v_{res}=
 c(1- {1 \over 2 \,  \gamma_{res}^2 } - {u_d^2\over 2  c^2}
) $ and $\epsilon =1 + 2 \delta$, $\delta \ll 1$.
From Eq.(\ref{aa11}) 
it is clear, that Cherenkov-drift resonance can be satisfied only
for superluminal particles with $ \gamma > 1/\sqrt{2 \delta}$.
  The emission geometry at  the Cherenkov-drift resonance is shown in
Figs. \ref{Cherenkov-driftemission}
and \ref{Cherenkovdrift1}.

Growth rate for the lt-mode in the limit $ K \approx 0$ is
\begin{equation}
\Gamma^{lt} = {4 \pi^2 q^2\over m } \int d p_{\phi}
\left({k_{\phi} k_x u_d  \over c k k_{\perp}} - {v_{\phi} \over c} {k_{\perp}
 \over k} \right)^2
{\partial f(p_{\phi}) \over  \partial p_{\phi} }
\delta\left(\omega- k_{\phi} v_{\phi} - k_x u_d \right),
\label{qaq7}
\end{equation}
and  growth rate for the t-mode is
\begin{equation}
\Gamma^{t} = {4 \pi^2 q^2\over m } \int d p_{\phi}
\left({k_r u_d  \over k_{\perp} c } \right)^2
 {\partial f(p_{\phi}) \over  \partial p_{\phi} }
\delta\left(\omega- k_{\phi} v_{\phi} - k_x u_d \right).
\label{qaq71}
\end{equation}

It is clear from (\ref{qaq7})
and (\ref{qaq71}) that the growth rate of the t-wave
is proportional to the drift velocity and becomes zero in the limit
of vanishing drift velocity.
  As for the lt-wave,  it can be excited in the limit
of vanishing drift by the conventional Cherenkov mechanism which does not
rely on the
  curvature of the magnetic field lines.
 We recall, that in the limit of
a strong magnetic field  and {\it oblique} propagation
 lt-wave has two branches: one superluminous and
one subluminous (\cite{arons1}).
On the conventional Cherenkov  resonance  it is
possible to excite only subluminous waves.

\section{Conclusion}
\label{concluss}

{ 
In this work we investigated 
electromagnetic processes associated with a 
  charged particle moving in a dielectric in a strong 
circular magnetic field. We derived a simple expression for
the growth rate of the Cherenkov-drift instability that, we
believe, may be responsible  for the generation of the pulsar 
radio emission.
We leave the discussion of the astrophysical applications of our
results for the subsequent paper (\cite{LyutikovBlandfordMachabeli}).
Here we just note, that the  Cherenkov-drift
developing on 
the open field lines in the outer parts of pulsar magnetosphere can 
explain  various features of the cone-type (\cite{rankin1}) emission patterns
observed in pulsars.

}

\acknowledgments
 ML would like to thank the Abastumani Astrophysics Observatory for
the hospitality during his stays in Tbilisi and GZM acknowledges the
support and  hospitality during his stay at Caltech.
This research was supported by NSF  grant AST-9529170.

{}

\newpage

\begin{figure}
\psfig{file=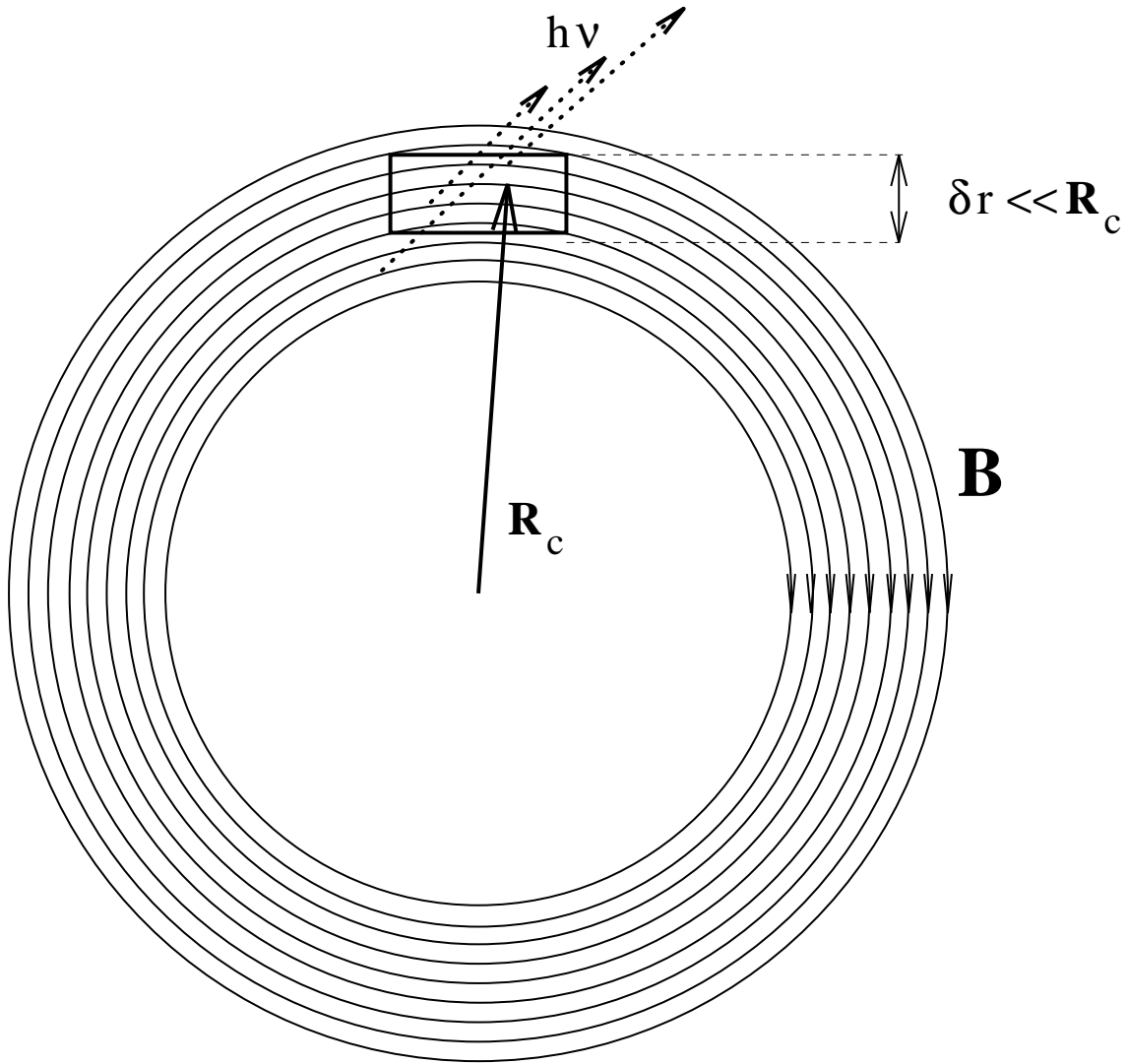,width=15.0cm}
\caption{
Geometry of  the considered problem. The magnetic field lines are
 concentric  coplanar
 circular arcs with the radius of
 curvature much larger than the size of the region.
\label{max1}}
\end{figure}

\begin{figure}
\psfig{file=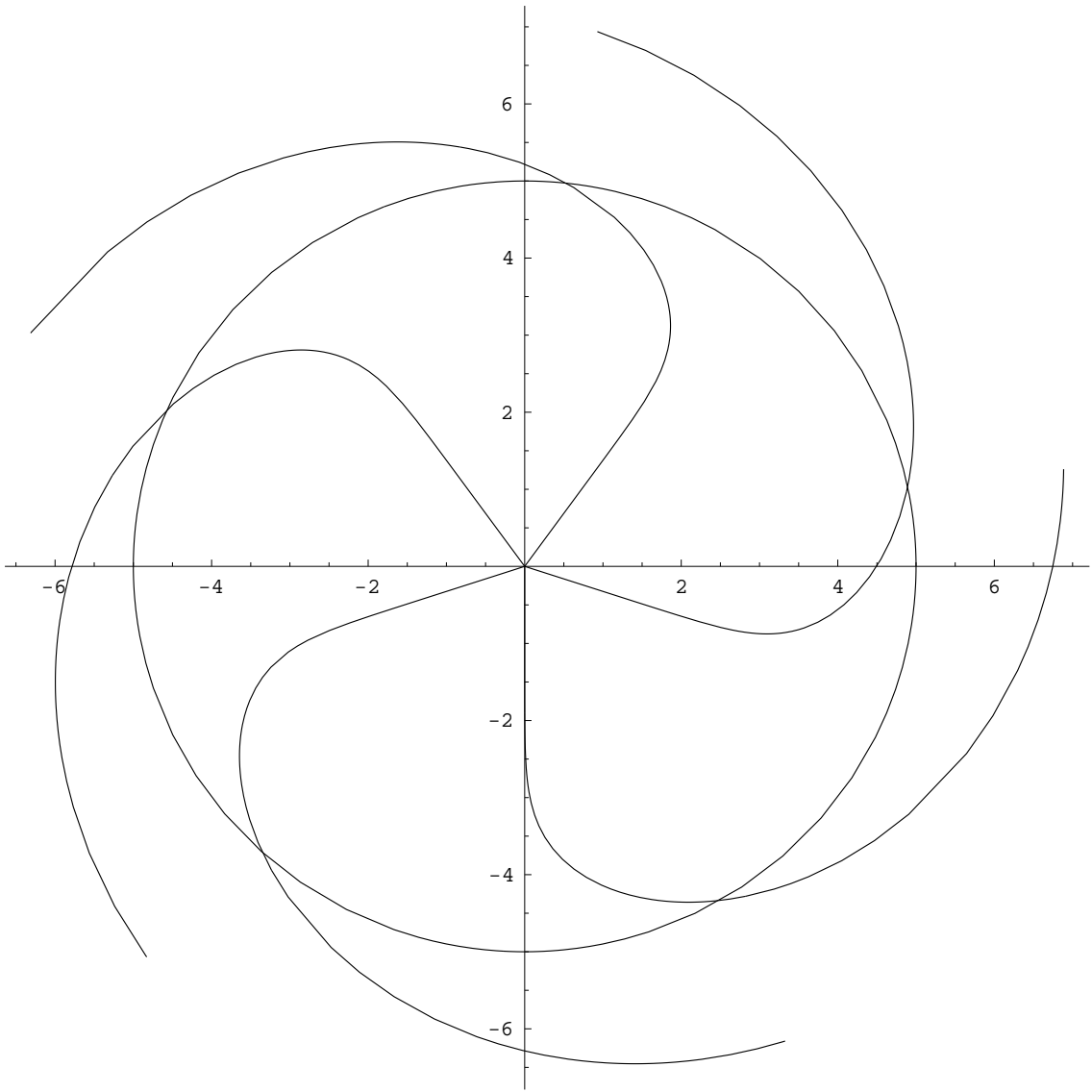,width=15.0cm}
\caption{
Surfaces of constant phase of  Hankel function $H_{\nu}^{(1)}$
for $\nu = 5$. Circle $\lambda r =\nu$  is an analog   of a light cylinder.
For radii much smaller than the "light cylinder" radius
$r_{\nu}= \nu/\lambda $, the rotation
is similar to solid body rotation, while for the radii much
larger  than the light cylinder radius
the surface of a constant phase has a form of unwinding   spiral with 
a wavelength $\lambda r/\nu$.
\label{phase}}
\end{figure}

\begin{figure}
\psfig{file=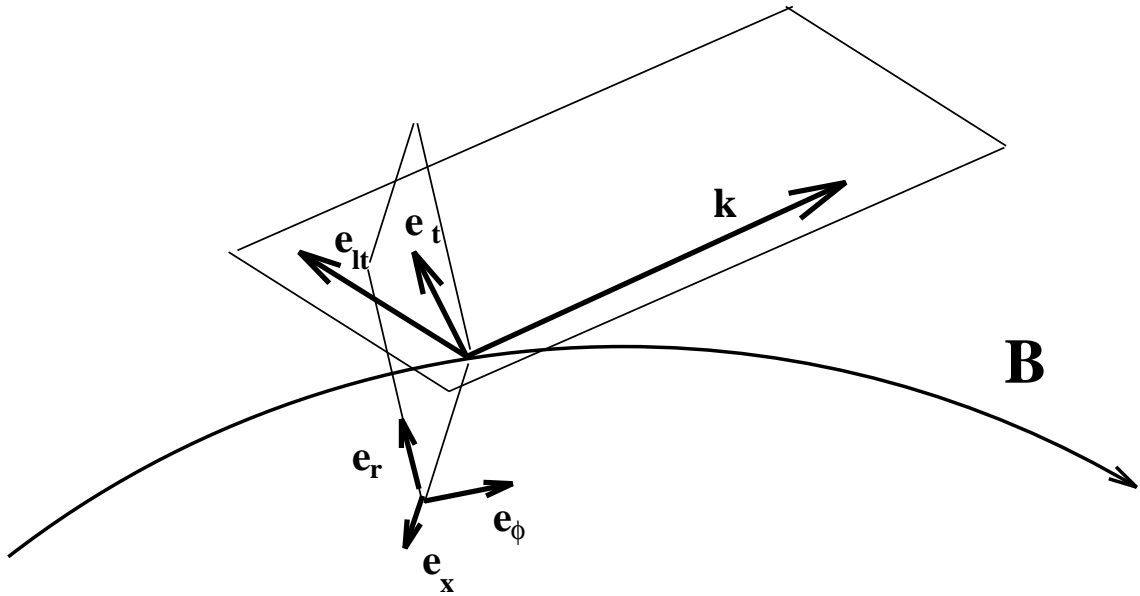,width=15.0cm}
\caption{
Polarization of normal modes
 in the limit of very strong magnetic field.
The electric field vector of the $t$-mode is in the plane
 ${\bf e_r} - {\bf e_x}$ and
the electric vector of the $lt $-mode is orthogonal to ${\bf e}_{t}-{\bf k}$ 
plane.
\label{Polariz}}
\end{figure}

\begin{figure}
\psfig{file=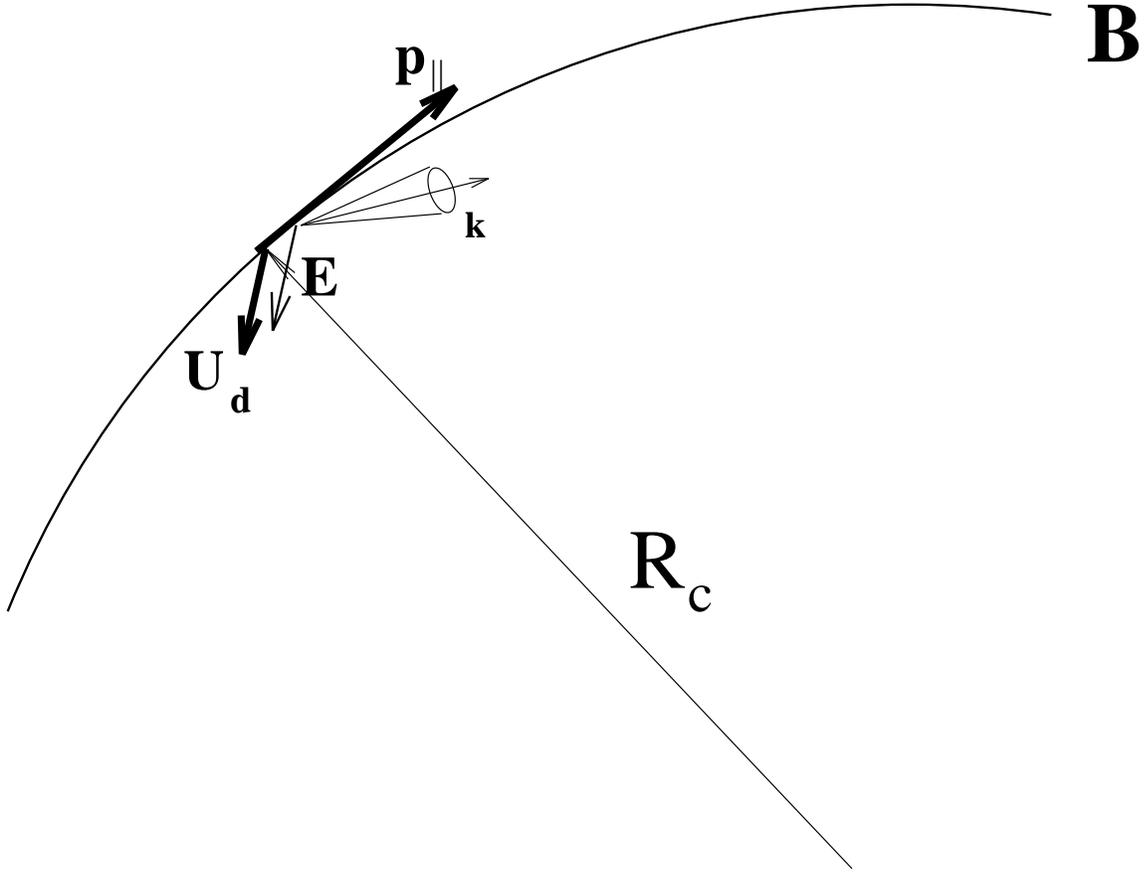,width=15.0cm}
\caption{
Cherenkov-drift emission in the case $\delta \ll u_d^2/c^2$.
 Drift velocity ${\bf u}_d$ is perpendicular
to the plane of the curved field line $({\bf B}-{\bf R_c} $
plane, ${\bf R_c} $ is a local radius of curvature).
The emitted electromagnetic waves
 are polarized along  ${\bf u}_d$. The emission is generated in the
 cone centered at the angle $\theta^{em} = u_d / c $ and
 with the opening angle  $(2 \delta)^{1/2} \ll \theta^{em}$.
\label{Cherenkov-driftemission}
}
\end{figure}

\begin{figure}
\psfig{file=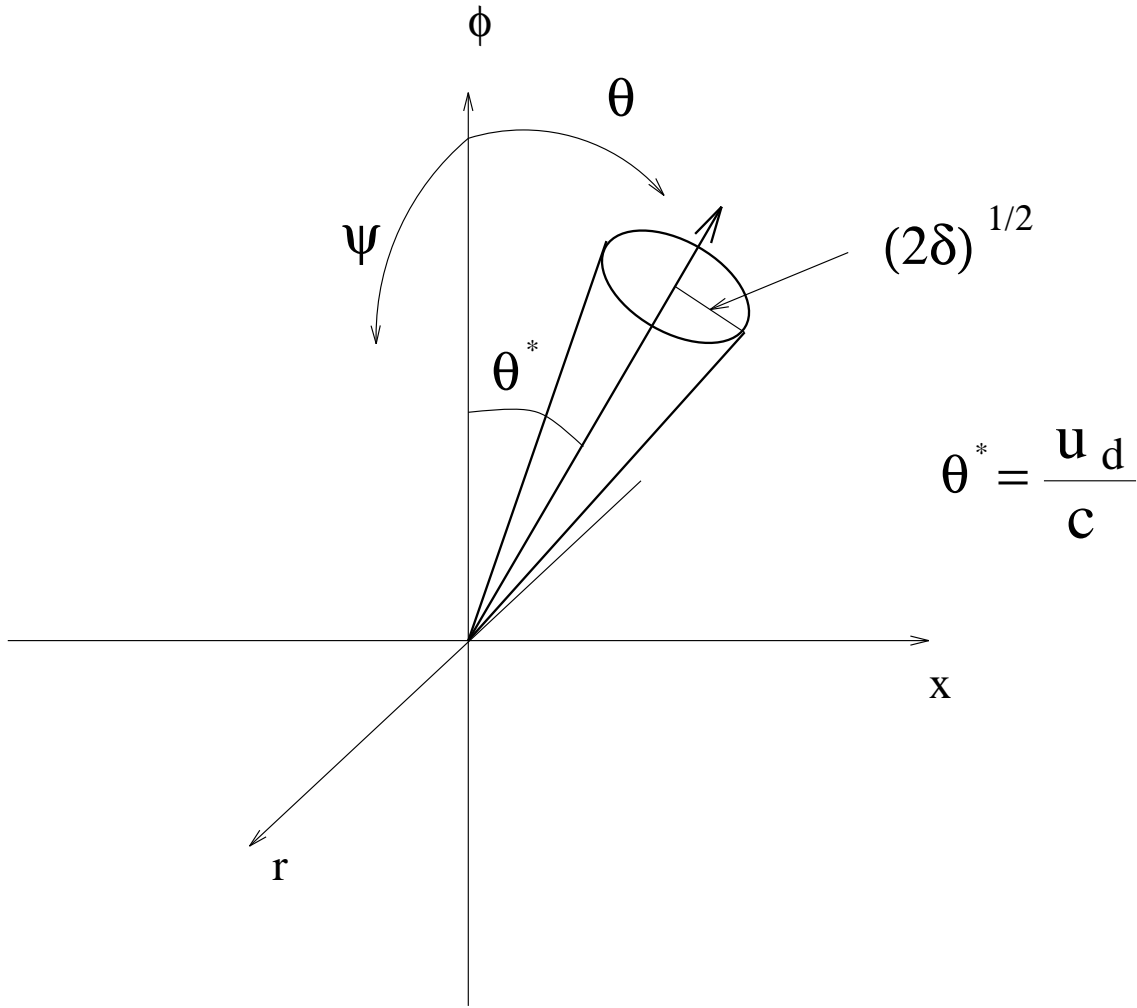,width=15.0cm}
\caption{
Emission geometry of the Cherenkov-drift resonance
\label{Cherenkovdrift1}}
\end{figure}

\end{document}